\documentclass[aps,reprint,prd,twocolumn,floatfix]{revtex4}%

\usepackage{amsfonts}
\usepackage{amsmath}
\usepackage{amssymb}
\usepackage{graphicx}
\usepackage[ansinew]{inputenc}
\usepackage{ulem}
\usepackage{pdfpages}
\usepackage{epsfig}
\usepackage{subfig}
\usepackage[colorlinks,hyperindex]{hyperref}%
\providecommand{\U}[1]{\protect\rule{.1in}{.1in}}
\hypersetup{
colorlinks,
citecolor=blue,
linkcolor=black,
urlcolor=black,
}

\begin{document}
\title{Quark stars with 2.6 $M_\odot$ in a non-minimal geometry-matter coupling theory of gravity}

\author{G. A. Carvalho$^a$, R. V. Lobato$^{b, c}$, D. Deb$^{d,e,\footnote{d.deb32@gmail.com}}$, P. H. R. S. Moraes$^{f}$ and M. Malheiro$^g$}
\affiliation{$^a$Departamento de F\'isica, Universidade Tecnol\'ogica Federal do Paran\'a, Medianeira, PR, Brazil \\
  $^b$Departamento de F\'isica, Universidad de los Andes,  Bogot\'a, Colombia PQ I-B processo 311327/2006-0\\
    $^c$ Departament of Physics and Astronomy, Texas A\&M University - Commerce, TX 75428, USA\\
    $^{d}$Department of Physics, Indian Institute of Science, Bangalore, 560012, India\\
    $^{e}$ The Institute of Mathematical Sciences, CIT Campus, Taramani, Chennai 600113, Tamil Nadu, India\\
$^{f}$Universidade Federal do ABC (UFABC) - Centro de Ci\^encias Naturais e Humanas (CCNH) - Avenida dos Estados 5001, 09210-580, Santo Andr\'e, SP, Brazil\\
$^g$Instituto Tecnol\'ogico de Aeron\'autica (ITA), 12228-900, S\~ao Jos\'e dos Campos, SP, Brazil
}

\keywords{Quark stars; non-minimal geometry-matter coupling; modified theories of gravity}

\begin{abstract}
    This work analyses the hydrostatic equilibrium configurations of strange stars in a non-minimal geometry-matter coupling (GMC) theory of gravity. Those stars are made of strange quark matter, whose distribution is governed by the MIT equation of state. The non-minimal GMC theory is described by the following gravitational action: $f(R,L)=R/2+L+\sigma RL$, where $R$ represents the curvature scalar, $L$ is the matter Lagrangian density, and $\sigma$ is the coupling parameter. When considering this theory, the strange stars become larger and more massive. In particular, when $\sigma=50$ km$^2$, the theory can achieve the 2.6 $M_\odot$, which is suitable for describing the pulsars PSR J2215+5135 and PSR J1614-2230, and the mass of the secondary object in the GW190814 event. The 2.6 $M_\odot$ is a value hardly achievable in General Relativity even considering fast rotation effects, and is also compatible with the mass of PSR J0952-0607 ($M = 2.35 \pm 0.17 ~M_\odot$), the heaviest and fastest pulsar in the disk of the Milky Way, recently measured, supporting the possible existence of strange quark matter in its composition. The non-minimal GMC theory can also give feasible results to describe the macroscopical features of strange star candidates.
\end{abstract}
\maketitle

\section{Introduction}\label{sec:int}

Recent observations regarding type Ia supernovae \cite{Perlmutter1999,Bennett2003,Riess1998} and cosmic microwave background radiation \cite{Spergel2003,Spergel2007,Ade2014} indicate that, presently, our universe is going through an accelerated expanding phase. Within the General Theory of Relativity (GR) context, the inclusion of the cosmological constant $\Lambda$ into the Einstein gravitational field equations is the standard way to explain cosmic acceleration and provide a good agreement with the observed data. However, the inclusion of $\Lambda$ faces a major setback due to a considerable mismatch of 120 orders of magnitude between its observational and theoretical values~\cite{Weinberg1989,Carroll2001}.

This situation engaged different researchers in more sophisticated gravity theories by modifying the Einstein-Hilbert action, which gave rise to a new avenue known as modified/extended gravity theories. These theories offer a great opportunity to solve problems that still do not have convincing explanations within the GR framework. In this regard, T. Harko and F.S.N. Lobo generalized the well-known $f(R)$-type gravity model \cite{amendola2007,hwang/2001} by assuming that the gravitational Lagrangian is given by an arbitrary function of the  Ricci scalar $R$ and the matter Lagrangian $L$, in the so-called $f(R,L)$ theory \cite{Harko2010}. The dynamics in such a theory can only exist in the presence of matter, which suggests a deeper link between spacetime and matter. In fact, the $f(R,L)$ gravity is a subclass of the geometry-matter coupling (GMC) theories \cite{wang/2018,nesseris/2009,delsate/2012}, i.e., theories that allow geometry and matter scalars to be mixed in the gravitational action.

The viability of $f(R,L)$ gravity as an alternative explanation for the cosmic acceleration was analyzed from a dynamical system approach in \cite{Azevedo2016}. Some constraints were obtained to $f(R,L)$ theories using the COBEFIRAS measurement of the cosmic microwave background spectral radiance \cite{Avelino2018}. The application of the energy conditions in the  $f(R,L)$ gravity can be seen in \cite{Wang2012, Wu2014}. Harko et al. have discussed the non-conservation of the energy-momentum tensor in $f(R,L)$ models in \cite{Harko2015, Harko2014D} and related it to a mechanism responsible for gravitationally induced particle production. Very recently, the $f(R,L)$ gravity was studied from a thermodynamic point of view \cite{pourhassan/2020}.

In reference \cite{Avelino2018D}, it was indicated that the $f(R,L)$ theories of gravity possibly be regarded as a subclass of the $f(R,T)$ gravity theories \cite{Harko2011}, in which $T$ is the trace of the energy-momentum tensor, with the latter theory also allowing for the GMC. In addition, in \cite{Harko2013} the $f(R,L)$ gravity action was generalized by inserting on it a scalar field. Moreover, a further model with GMC was proposed by Harko in  \cite{Harko2008}. In \cite{Moraes2017, Harko2010D} it was shown that GMC models can be candidates to solve fundamental issues of standard gravity, without considering dark energy \cite{Copeland2006,Frieman2008} and dark matter \cite{navarro/1996,moore/1999,bertone/2005}. Note that it is generally believed that dark energy is the cause of accelerated expansion. On the other hand,  dark matter is an exotic matter that does not interact with light but interacts gravitationally and strongly affects the galactic and intergalactic dynamics. For a review of generalized GMC theories, one can also check \cite{Harko2014E}.

A well-behaved extended or alternative theory of gravity must also show a significant effect on the stellar astrophysics regime. In other words, a given alternative theory of gravity should exhibit substantial effects on the cosmological and galactic scales, as well as predict the existence of observable stable, compact stellar objects, such as white dwarfs, neutron stars and black holes. In fact, the study and analysis of compact objects are of great importance in astrophysics because these objects provide an excellent laboratory to study dense matter in extreme conditions, such as the strong gravity regime. In particular, neutron stars were already studied in $f(R,L)$ gravity, providing a remarkable increase in the maximum mass limit \cite{Carvalho2020May,Lobato2021Nov}. In \cite{Carvalho2020May}, the matter inside neutron stars was described by a relativistic polytropic equation of state (EoS) and also a Skyrme type EoS known as SLy4. It was shown in this theory that the mass of massive pulsars can be achieved, such as PSR J2215 + 5135, for both equations of state. It was pointed out that results for mass-radius relation in GMC gravity strongly depend on the stiffness of the EoS. In this theory, a further investigation of neutron stars considering several realistic nuclear matter equations of state was performed in \cite{Lobato2021Nov}. NS masses and radii obtained were subject to observational constraints from massive pulsars, the gravitational wave event GW170817 and the PSR J0030+0451 mass-radius from NASA's Neutron Star Interior Composition Explorer (NICER) data. It was shown that in this theory of gravity, the mass-radius results could accommodate massive pulsars. The mass-radius relation in GMC theory shows a sharp increase in the star mass around one solar mass up to 2.25 $M_{\odot}$ with a very small change in the star radius, which is $\sim 13$ Km for a strong gravity matter coupling. These results agree with the NS radius region constrained by PSR J0030+0451 and by the very massive PSR J0740+6620 obtained in NICER observations \cite{Riley2019Dec,Miller2021Sep,Riley2021Sep}, and also in accordance with the
GW170817 event \cite{abbott/2017}. Very recently, the mass of the pulsar PSR J0952-0607, the fastest known spinning neutron star (NS) in the disk of the Milky Way, has been measured \cite{Romani2022Jul}. This pulsar was firstly reported in \cite{Bassa2017Sep} and it has a spin period of $P = 1.41$ ms. It is a ``black widow" pulsar with a low-mass (substellar) companion being irradiated and evaporated by the pulsar's luminosity.  The mass measurement of PSR J0952-0607 indicates a maximal mass of 2.52 $ M_\odot$ ($M = 2.35 \pm 0.17 M_\odot$ \cite{Romani2022Jul}). This high mass value presents for sure the most severe constraint on the dense-matter equation of state, and can also indicate the existence of exotic matter in its interior, a possibility that we will explore in this work.

The hydrostatic equilibrium configurations of quark stars in a non-minimal GMC model have not been investigated. The strange quark matter, made up of approximately equal numbers of unconfined up, down and strange quarks, may be the absolute ground state of the strong interaction \cite{Farhi1984,Alcock1988,Bodmer1971,Witten1984}. There may exist objects entirely made of strange matter \cite{Malheiro2003Apr,moraes/2014}. Note that three flavor strange quark matter is more stable than the two flavor non-strange ones \cite{Weber2005, Panotopoulos2019}. As a result, a new class of compact objects has been postulated to exist, namely quark stars, almost self-bounded systems with an energy density at the star surface - given in the MIT model by the confined Bag energy $\mathcal{B}$ - in contrast with NSs that are bounded by gravity. Some constraints have been put to quark stars from gravitational waves \cite{zhou/2018,gomes/2019}, particularly from the GW170817 event \cite{abbott/2017}. 

In the present work, we are particularly interested in analyzing quark stars in a non-minimal GMC model, which shall be presented in Section \ref{secII}. In Section \ref{secIII} the hydrostatic equilibrium equations for the concerned theory will be reviewed. In Section \ref{secIV} we will present the equation of state (EoS) that we shall consider for numerically solving the hydrostatic equilibrium equations in Section \ref{secV}. In Section \ref{secVI} we present our conclusions.

\section{Basic formalism}
\subsection{A Non-minimal Geometry-Matter Coupling Theory}\label{secII}

The concerned modified form of the Einstein-Hilbert action  reads \cite{Harko2010}:
\begin{equation}\label{action}
    S= \int d^4 x \sqrt{-g}f(R,L),
\end{equation}
being $f(R,L)$ an arbitrary function of $R$ and $L$. The constants $8\pi G$ and $c$, with $G$ being the Newtonian gravitational constant and $c$ the speed of light, are taken as $1$. One can observe from Eq. \eqref{action} that when $f(R,L)=R/2+L$, the standard form of Einstein-Hilbert action is retrieved, which leads to the standard Einstein's field equations: $G_{\mu\nu}=T_{\mu\nu}$, where $T_{\mu\nu}$ is the energy-momentum tensor.

Following \cite{Garcia2010, Garcia2011}, we consider a GMC model defined by $f(R,L)=R/2+L(1+\sigma R)$. Furthermore, we will assume $L=-p$ \cite{Harko2014}, where $p$ is the pressure of the fluid. Applying the variational principle in \eqref{action}, and taking into account the considerations above, it follows that the field equations become \cite{Harko2010,Carvalho2020May}
\begin{eqnarray}\label{eq02}
(1-2 \sigma p) G_{\mu \nu}+\frac{1}{3} R g_{\mu \nu}-\frac{\sigma p}{3} R g_{\mu \nu}= \nonumber\\ (1+\sigma R)\left(T_{\mu \nu}-\frac{1}{3} T g_{\mu \nu}\right)-2 \sigma \nabla_{\mu} \nabla_{\nu} p.
\end{eqnarray}

Furthermore, the covariant derivative of the energy-momentum tensor reads \cite{Harko2010,Carvalho2020May}
\begin{equation} \label{eq03}
\nabla^{\mu} T_{\mu \nu}=\left(-p g_{\mu \nu}-T_{\mu \nu}\right) \nabla^{\mu} \ln (1+ \sigma R).
\end{equation}

\subsection{The Hydrostatic Equilibrium Equations in a Non-minimal Geometry-Matter Coupling Model}\label{secIII}

The hydrostatic equilibrium equations in the concerned GMC theory were previously derived by adopting a spherically symmetric metric in its canonical form,
\begin{equation}\label{ds2}
d s^{2}=e^{\alpha(r)} d t^{2}-e^{\beta(r)} d r^{2}-r^{2}\left(d \theta^{2}+\sin ^{2} \theta d \phi^{2}\right),
\end{equation}
with $\alpha(r)$ and $\beta(r)$ being the potentials depending on $r$ only. The energy-momentum tensor of a perfect fluid is defined as $T_{\mu\nu}={\rm diag}(e^{\alpha}\rho,e^{\beta}p,r^2 p,r^2\sin^2\theta p)$, where $\rho$ is the matter-energy density. From the substitution of \eqref{ds2} into \eqref{eq03} one can find that the energy-momentum tensor is covariantly conserved independently of the functional form assumed for $f(R,L)$. Detailed derivation is presented in \cite{Carvalho2020May}.

The equilibrium configurations are obtained from the 00 and 11 components of the field equations,
\begin{eqnarray}\label{eq:2}
\frac{(1-2 \sigma p)}{r^{2}}\left[r(1- e^{-\beta})\right]^{\prime}+(1-\sigma p) \frac{R}{3}\nonumber \\ =(1+\sigma R)\left(\frac{2}{3} \rho+p\right)+\sigma e^{-\beta} \alpha^{\prime} p^{\prime},
\end{eqnarray}
\begin{eqnarray}\label{eq:3}
\frac{(1-2 \sigma p)}{r^{2}}\left[e^{-\beta}(1- \alpha^{\prime} r)-1\right]+(\sigma p-1) \frac{R}{3} \nonumber\\ =(1+\sigma R) \frac{\rho}{3}-2 \sigma e^{-\beta}\left(p^{\prime \prime}-\frac{\beta^{\prime}}{2} p^{\prime}\right),
\end{eqnarray}
where prime represents the derivative with respect to radial coordinate $r$.

The first term between parenthesis on the right-hand side of \eqref{eq03} vanishes given the assumed metric and matter Lagrangian, which leads to the covariant conservation of the energy-momentum tensor. Another way to a conservative energy-momentum tensor is to have $\sigma R<< 1$.

\begin{equation}\label{eq04}
p^{\prime}=-(\rho+p) \frac{\alpha^{\prime}}{2}.
\end{equation}

Another equation can be derived from the trace of the field equations and reads
\begin{equation}\label{eq05}
(1+2 \sigma p) R=-(1+\sigma R) T-6 \sigma \square p,
\end{equation}
where $\Box$ is the D'Alambertian operator, defined as
\begin{equation}
\square=-e^{-\beta}\left(\frac{d^{2}}{d r^{2}}-\frac{\beta^{\prime}}{2} \frac{d}{d r}+\frac{\alpha^{\prime}}{2} \frac{d}{d r}+\frac{2}{r} \frac{d}{d r}\right).
\end{equation}

In order to obtain the hydrostatic equilibrium configurations, \eqref{eq05} must be included in the system of differential equations \eqref{eq:2}, \eqref{eq:3} and \eqref{eq04}. The unknowns are $R, \alpha, \beta, \rho$ and $p$. An EoS must be provided to solve the set of differential equations. Details about the numerical procedure to solve the system of equations \eqref{eq:2}, \eqref{eq:3}, \eqref{eq04} and \eqref{eq05} are given in \cite{Carvalho2020May}. 

In previous works on compact stars in geometry-matter coupling theories of gravity \cite{Carvalho2020May,Lobato2021Nov,Lobato2022Jun}, authors have used the following formulation to determine the stellar mass 
\begin{equation}\label{mass-calc}
    M=\int_0^{R_\star} 4\pi r^2 \rho {\rm d}r,
\end{equation}
where $R_\star$ represents the stellar radius, which is by definition the point where pressure vanishes, $p(R_\star)=0$. However, Eq. \eqref{mass-calc} gives the baryonic mass of the star. In the absence of matter (vacuum), the matter Lagrangian $L$ vanishes, and the new terms arising from the GMC theory disappear, so, from $R_\star$ forward the metric must obey the exterior Schwarzschild solution. The gravitational mass may differ from the baryonic one due to additional contributions from the theory. To obtain the gravitational mass one can make use of the junction condition, i.e.,
\begin{equation}\label{gravitmass}
    e^{-\beta(R_\star)}=1-\frac{2M}{R_\star},
\end{equation}
which means that at the surface the interior solution must smoothly connect to the exterior one. Hence, from \eqref{gravitmass} one can calculate the gravitational mass (perceived by a distant observer) as
\begin{equation}\label{gravitmass2}
    M=\frac{(1-{\rm e}^{-\beta(R_\star)})R_\star}{2}.
\end{equation}
In particular, Eq. \eqref{gravitmass2} gives smaller values for the mass in comparison with \eqref{mass-calc} for a given central energy density $\rho_C$, which indicates that the gravitational mass has negative contributions from the GMC theory, thus becoming smaller than the baryonic one.

\section{Equation of State for Nuclear Matter inside Quark Stars\label{secIV}}

The EoS for the matter inside the star is considered to be that for quark-gluon plasma, dubbed MIT bag model \cite{Farhi1984Dec}. To consider hadronic masses in terms of their constituents, the MIT bag model considers that the quarks are inside a ``bag'' which reproduces the asymptotic freedom and confinement, i.e.,
\begin{equation}
    p = \omega (\rho -4 \mathcal{B}),
\end{equation}
where $\omega$ is the EoS parameter and $\mathcal{B}$ is the bag constant, which we take as $\mathcal{B}=60$ MeV/fm$^3$ in allusion to \cite{moraes/2016}, among many others. We adopt this bag constant value since the parameters of the maximum mass configurations for quark stars are similar to those for realistic neutron stars made of baryonic matter. The value of $\omega$ is related to the Quantum Chromodynamics coupling constant, and the strange quark mass \cite{Jaffe1979Apr}. For $\omega=0.28$, the mass of the strange quark is $m_s=250$ MeV, and $\omega=1/3$ for massless quarks. In this work, we adopt $\omega=0.28$.

\begin{table*}
    \caption{\label{tab2} Physical parameters of observed strange star candidates derived using $\sigma=30$ km$^2$ and $B=60$ MeV/fm$^3$.}
    \begin{tabular}{lcccccr} 
        \hline
      SS candidate & Observed mass $M/M_{\odot}$ & Predicted radius (km) & $M/R_\star$ & $Z_s$ & $\rho_C$ (MeV/fm$^3$) & $p_C$ (MeV/fm$^3$) \\
        \hline
       PSR J2215 + 5135 & $2.27^{+0.17}_{-0.15}$ \cite{linares/2018} &  $10.74^{+1.28}_{-0.94}$ & 0.312  & 0.631  & 2760 & 705  \\   
       PSR J1614-2230  & 1.97$\pm$0.04 \cite{Demorest2010Oct}  & $12.22_{+0.01}$ & 0.238  & 0.382 & 711 & 132\\
       Vela X-1  & 1.77$\pm$0.08 \cite{Rawls2011Feb}  & $12.08^{+0.08}_{-0.06}$ & 0.216  & 0.328 & 562 & 90\\
       4U 1608-52  & 1.74$\pm$0.14 \cite{Guver2010Mar}  & $12.06^{+0.12}_{-0.16}$ & 0.213  & 0.320 & 546 & 86\\
       PSR J1903+327 & 1.667$\pm$0.021 \cite{Freire2011Apr} & $11.99^{+0.004}_{-0.01}$ & 0.205  & 0.303 & 510 & 76\\
       4U 1820-30  & 1.58$\pm$0.06 \cite{Guver2010Aug} & $11.87{\pm 0.09}$ & 0.197  & 0.284 & 478 & 67\\
       Cen X-3  & 1.49$\pm$0.08 \cite{Rawls2011Feb} & $11.72\pm 0.13$ & 0.188  & 0.265 & 451 & 59\\
       EXO 1785-248  & 1.3$\pm$0.2 \cite{Ozel2009Mar} & $11.36^{+0.4}_{-0.5}$ & 0.169 & 0.229 & 402 & 45\\
       LMC X-4  & 1.29$\pm$0.05 \cite{Rawls2011Feb} & $11.34\pm 0.1$ & 0.168  & 0.227 & 400 & 45\\
       SMC X-1  & 1.04$\pm$0.09 \cite{Rawls2011Feb} & $10.71\pm 0.2$ & 0.143 & 0.184 & 356 & 32\\
       SAX J1808.4-3658  & 0.9$\pm$0.3 \cite{Elebert2009Apr} & $10.28^{+0.8}_{-1.1}$ & 0.129  & 0.161 & 336 & 27\\
       4U 1538-52  & 0.87$\pm$0.07 \cite{Rawls2011Feb}  & $10.17\pm 0.2$ & 0.126 & 0.157 & 331 & 28\\
       HER X-1  & 0.85$\pm$0.15 \cite{Abubekerov2008May}  & $10.11\pm 0.5$ & 0.124  & 0.153 & 329 & 25\\
       \hline
    \end{tabular}
\end{table*}

\begin{table*}
         \caption{\label{tab1} Maximum masses with their correspondent radii, central energy densities and central pressures for some values of $\sigma$.}
    \begin{tabular}{lcccr} 
        \hline
      $\sigma$ (km$^2$) & Maximum mass ($M/M_\odot)$ & Radius (km) & $\rho_C$ (MeV/fm$^3$) & $p_C$ (MeV/fm$^3$)\\
        \hline
       0 & 1.76 & 10.19 & 1284 & 292\\    
       10 & 1.88 & 10.62  & 1370 & 316\\
       \hline
    \end{tabular}
\end{table*}

\begin{table*}
         \caption{\label{tab3} Physical parameters of the strange star candidate LMC X-4 for different values of $\sigma$ and $B=60$ MeV/fm$^3$.}
    \begin{tabular}{lccccr} 
        \hline
      $\sigma$ (km$^2$) & Predicted radius (km) & $M/R_\star$ & $Z_s$ & $\rho_C$ (MeV/fm$^3$) & $p_C$ (MeV/fm$^3$)\\
        \hline
       0 & 10.38 & 0.184 & 0.257 & 454 & 60\\    
       10 & $10.74$ & 0.178  & 0.246 & 432 & 53\\
       30  & $11.34$ & 0.168 & 0.227 & 400 & 45\\
       50 & 11.91 & 0.160 & 0.213 &  379 & 39\\
       \hline
    \end{tabular}
\end{table*}

\section{Results}\label{secV}

\begin{figure}
  \centering
  \includegraphics[scale=0.52]{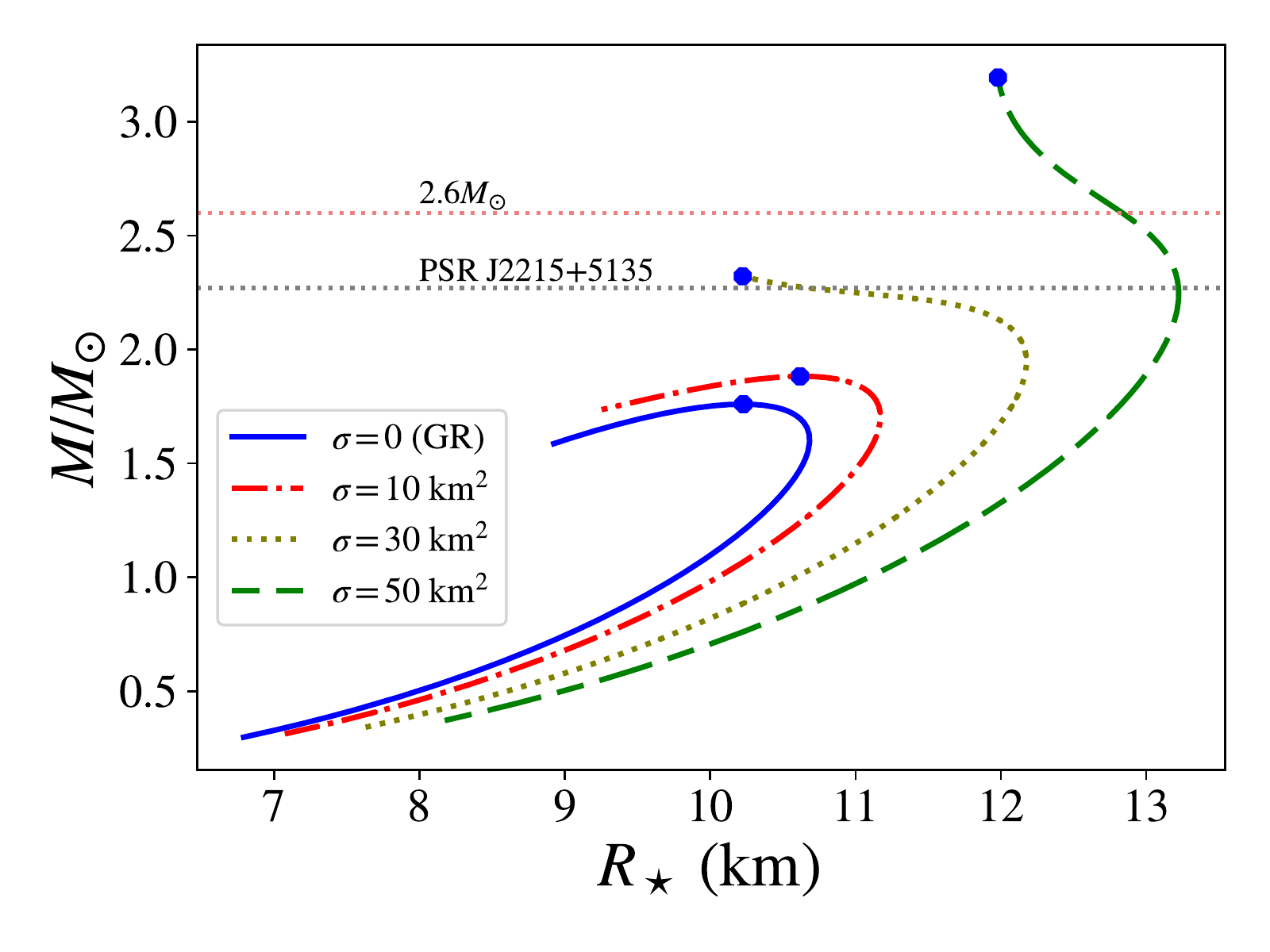}
  \caption{Mass-radius relation of quark stars for different values of $\sigma$.}\label{mass-radius}
\end{figure}

\begin{figure}
  \centering
  \includegraphics[scale=0.52]{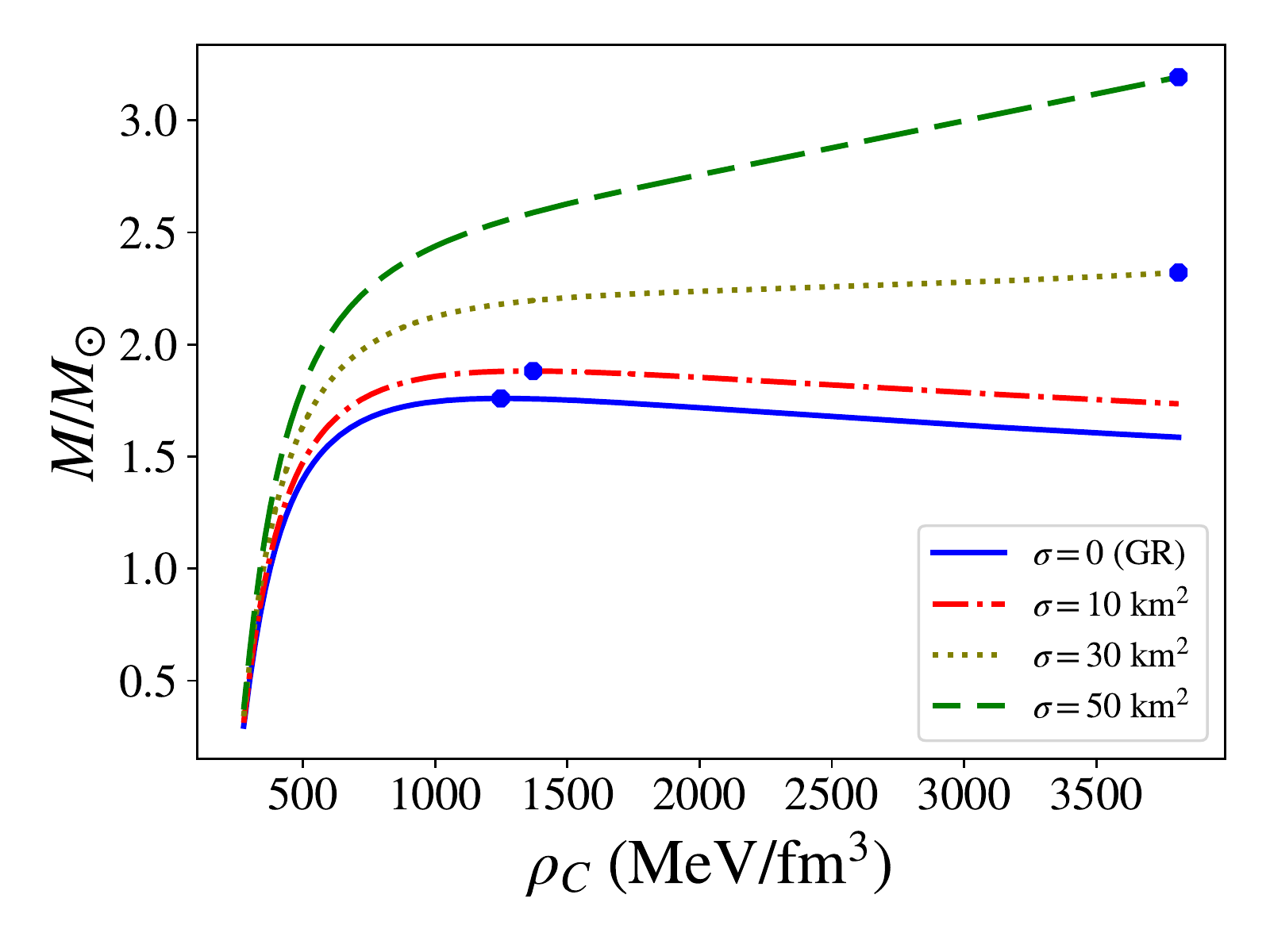}
  \caption{Mass-central energy density relation of quark stars for different values of $\sigma$.}\label{mass-rho}
\end{figure}

In Figure \ref{mass-radius}, we present the mass-radius relation for quark stars in the GMC theory of gravitation for four values of $\sigma$, the GMC parameter. Blue circles mark the maximum mass points for each value of $\sigma$. The two horizontal lines represent the values of mass 2.3 and 2.6 $M_{\odot}$, which correspond to the mass of PSR J2215+5135 \cite{linares/2018}, and the mass of the lighter object in the GW190814 event \cite{abbott/2020}, respectively. When $\sigma$ equals zero, GR results are retrieved. From the figure, it is possible to see that the radii of the stars range from $7$ to $13$ km approximately. This radius interval lies within the expected values of compact star radii constraints from observed X-ray binaries, and GW170817 gravitational wave event \cite{Most2018Jun,Marino2018Jun}. 

Figure \ref{mass-radius} also shows that maximum mass points change according to the value of $\sigma$. For larger $\sigma$, more massive stars are found. The radii corresponding to the maximum masses also increase with increasing $\sigma$. This is similar to the effects of charge on the stellar structure of compact objects \cite{Liu2014May,Arbanil2015Oct,Carvalho2018May,rocha/2020}, where increasing total charge increases the minimum radius. It is important to stress that in the GMC theory, the maximum star mass is not only a function of the bag constant but also depends on $\sigma$, while in GR, the quark star mass increases with the star volume almost up to the maximum star mass that is only a function of the bag constant \cite{Banerjee2000Jan,Cheng2000Sep,Harko2002Apr}.

A remarkable feature of the GMC theory is that large values for the coupling parameter (30 km$^2$ and 50 km$^2$, for instance) do not produce a region with ${\rm d}M/{\rm d}R>0$. {\it A priori}, for these values of $\sigma$ the theory does not have a maximum mass limit, since it does not have a maximum mass point. However, the mass will be constrained by Buchdahl, Schwarzschild and causal limits. Increasing $\sigma$ also leads to increased compactness. From Figure \ref{mass-radius}, it is clear that the maximum mass stars have radii between 9-12 km, indicating that the Buchdahl and Schwarzschild limits are respected, so the mass will be restricted by causality.

Figure \ref{mass-rho} presents the mass-central density relation for the same four values of $\sigma$ as in Figure \ref{mass-radius}. By increasing $\sigma$, maximum masses are always found for larger central energy densities. At smaller densities, the GMC results become closer to GR ones ($\sigma=0$). This indicates that GMC effects are more evident at a high density regime, so, in less compact systems, such as non-compact stellar objects or in the solar system regime, GMC outcomes would be the same as the GR ones. Table \ref{tab1} below is a list of the stellar parameters of the maximum mass quark stars in GMC theory, where we can see that the central energy densities of the maximum mass stars are between 2-20 times the nuclear saturation energy density ($\rho_0 \sim 140 $ MeV/fm$^3$).

We see that GMC theory can enhance maximum masses, allowing compact stars to sustain more mass against gravitational collapse, which indicates that the theory is capable of describing the data of massive compact stars, such as PSR J2215+5135 \cite{linares/2018} and PSR J1614-2230 \cite{Demorest2010Oct}. 

Recently, the LIGO (Laser Interferometer Gravitational wave Observatory)/VIRGO experiments have  detected the GW190814 event \cite{abbott/2020}, which is a binary merger of two compact objects, with one of them being possibly a neutron star or a strange quark star with a mass of 2.6$M_\odot$. This mass value can be attained with $\sigma\approx 40$, which sets an observational upper limit to the GMC parameter.

\begin{figure}[]
  \centering
  \includegraphics[scale=0.52]{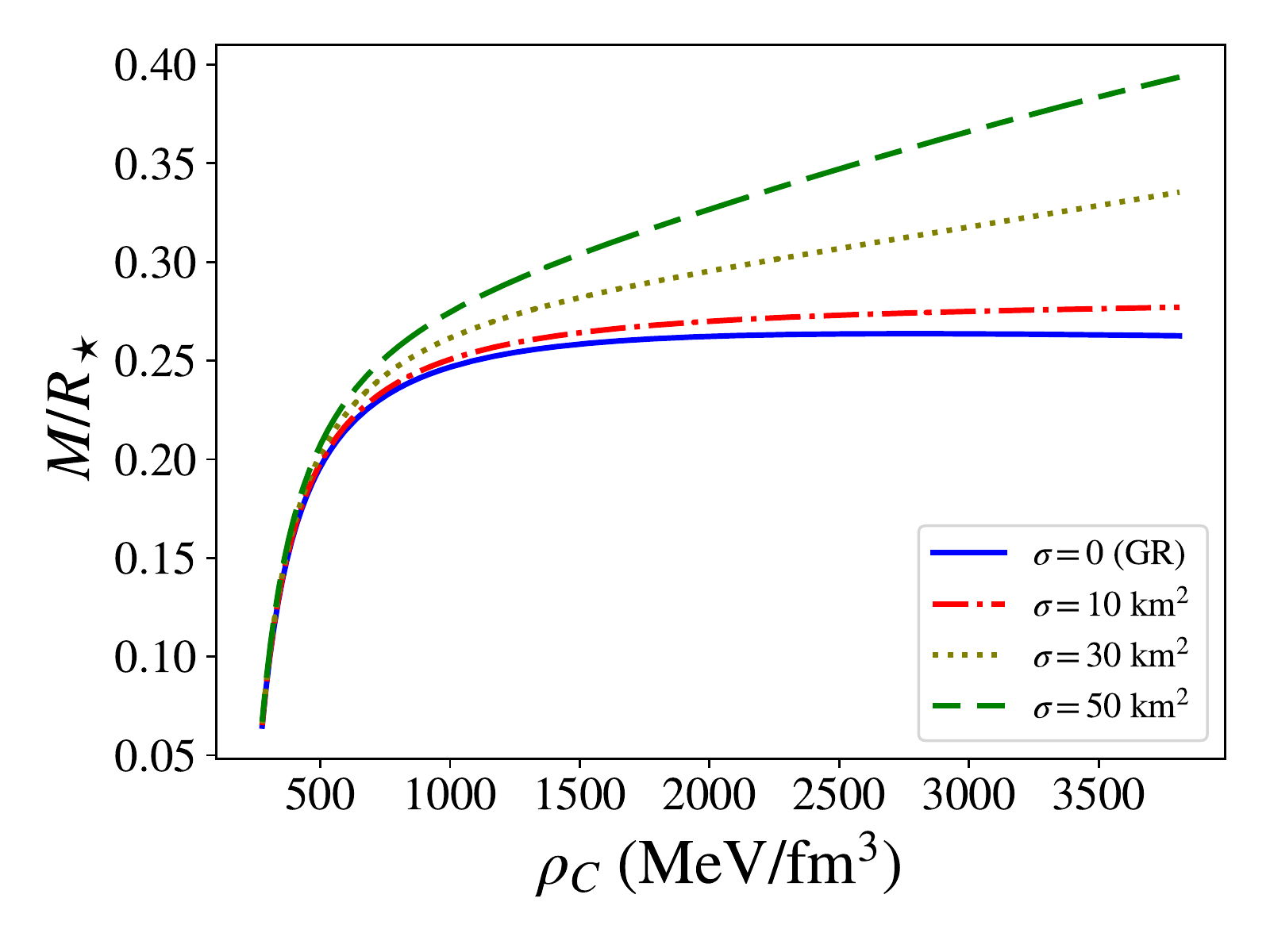}
  \includegraphics[scale=0.52]{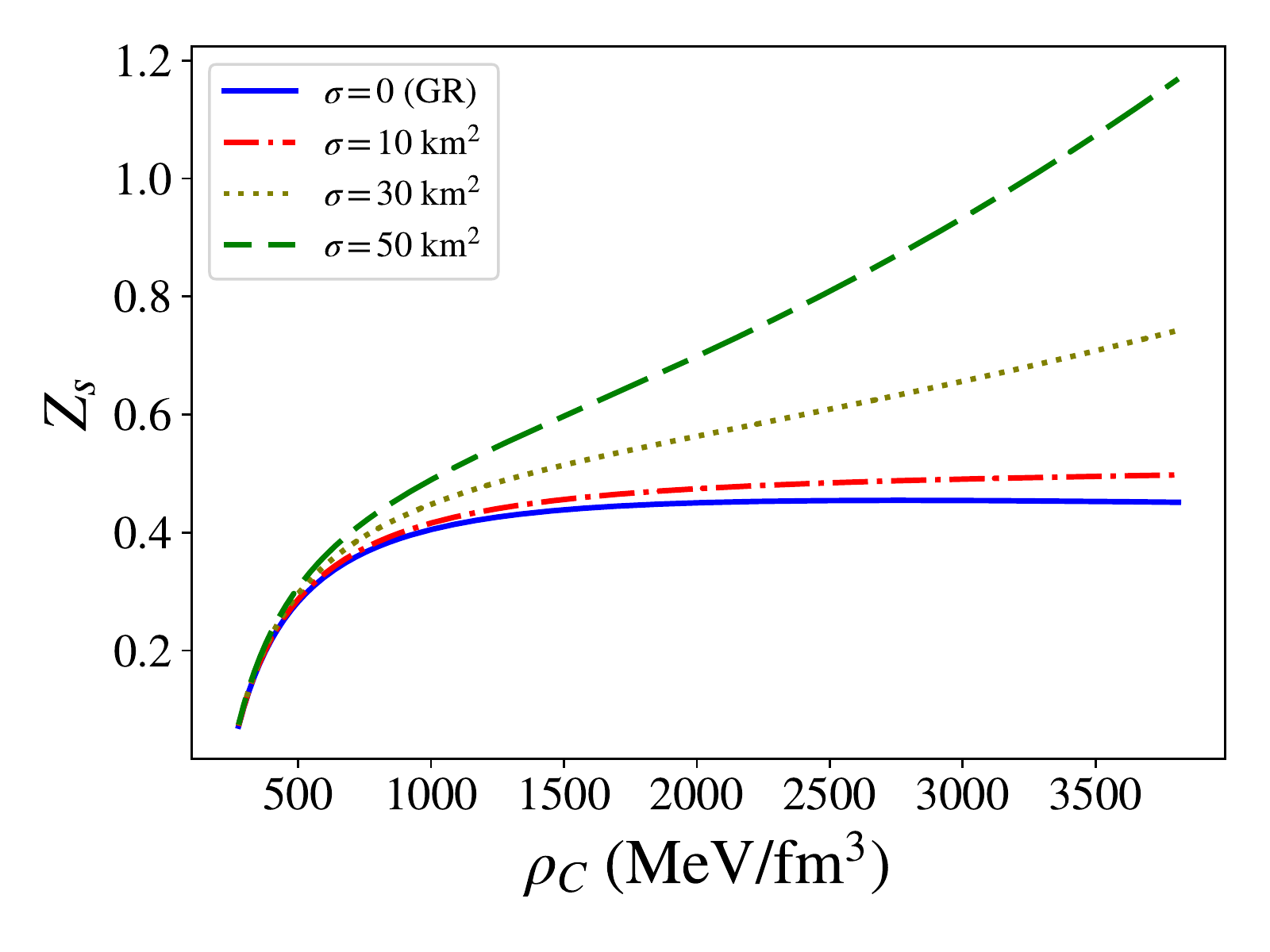}
  \caption{\label{compc-reds} Compactness and surface red-shift as function of the star's central energy density.}
\end{figure}

In Figure \ref{compc-reds}, we present the compactness and surface redshift as a function of the central energy density of the stars. These quantities are given by
\begin{equation}
    u=\frac{M}{R_\star} \qquad {\rm and} \qquad Z_s= \frac{1}{\sqrt{1-2M/R_\star}}-1,
\end{equation}
respectively. It is observed that as the parameter $\sigma$ increases, the compactification degree is higher. The surface redshift is also shown to increase with central energy density and with $\sigma$. 

\begin{figure}[t]
  \centering
  \includegraphics[scale=0.52]{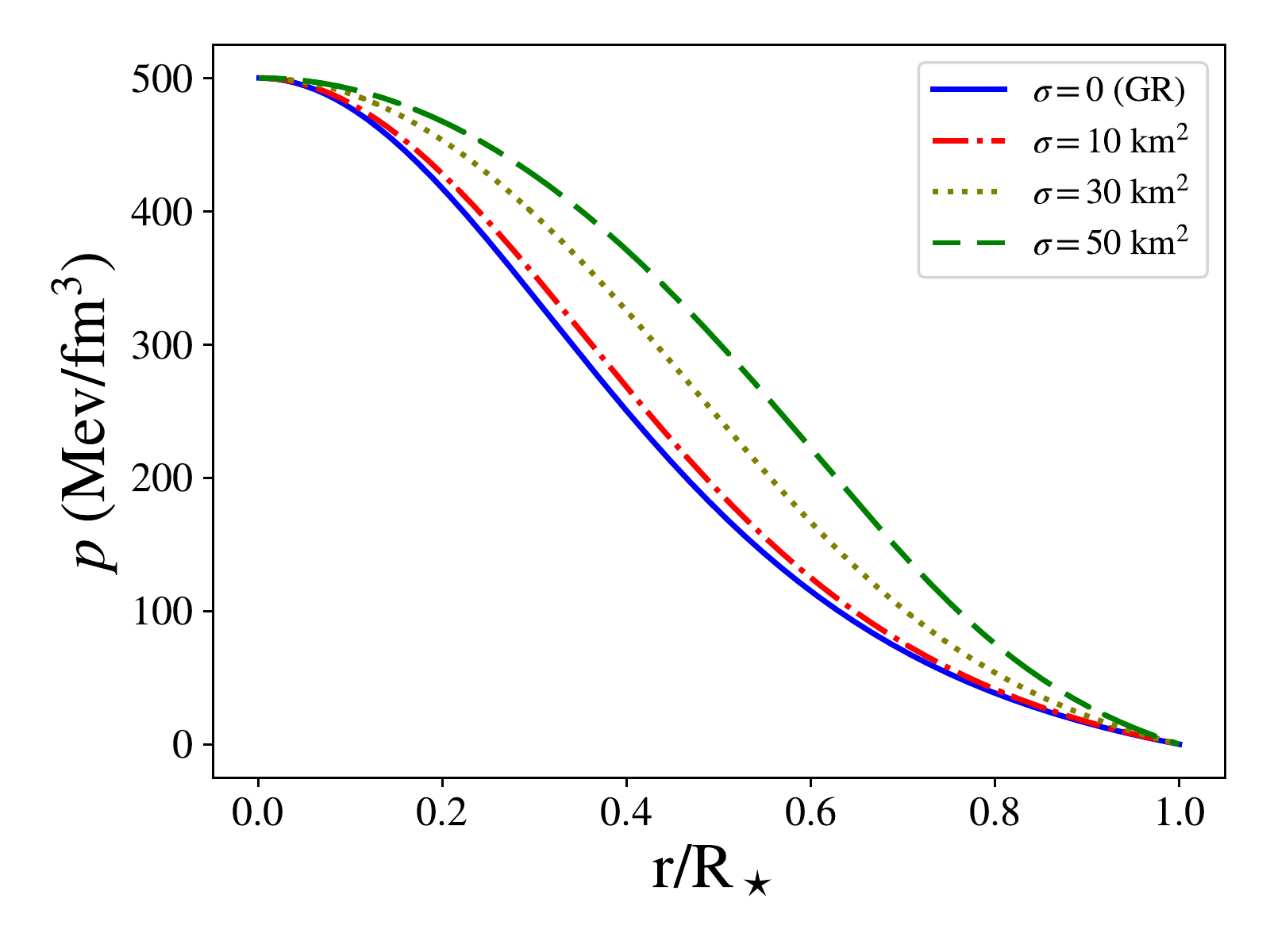}
  \includegraphics[scale=0.52]{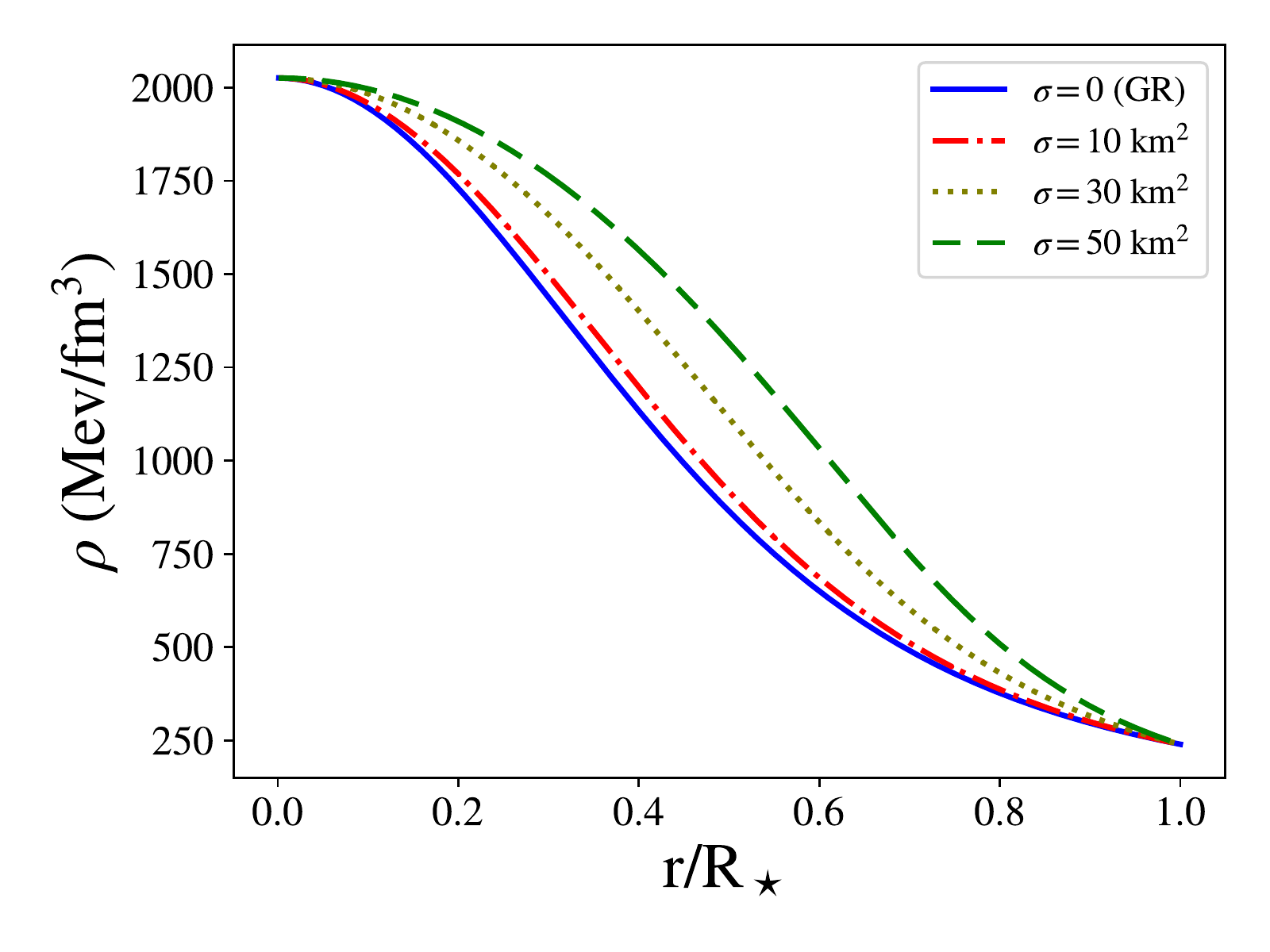}
  \caption{\label{profs} Profiles of pressure and energy density inside a quark star with central pressure of 500 MeV/fm$^3$. This is an example to illustrate that the GMC theory yields to well-behaved scenarios, where it is straightforward to show that energy conditions are respected.}
\end{figure}

In Figure \ref{profs}, we show the energy density and pressure profiles for a strange quark star with a central pressure of 500 MeV/fm$^3$ in the GMC theory. Pressure and energy density are shown to decrease as radial coordinate increases. In the cases where $\sigma\neq 0$, density and pressure are larger and take longer to diminish. This yields a larger mass according to increasing $\sigma$ and a slightly larger radius. Moreover, the variations of pressure and energy density are always negative, indicating that the energy conditions \cite{morris/1988} are respected.

In Table \ref{tab2}, we have used $\sigma=30$ km$^2$ and $B=60$ MeV/fm$^3$ to derive the radius, compactness, surface redshift, central energy density and the central pressure of strange star candidates. The table shows that the stars have high surface redshift (0.15-0.63) and compactness (0.12-0.31), which is reinforced by the results for central energy densities (2-20 $\rho_0$, where $\rho_0$ represents an average value for nuclear saturation energy density). From Figure \ref{mass-radius} and Table \ref{tab2}, one can observe that most stars have radii within the range of 10-11 km, which can be a good marker to analyze strange quark stars in GMC theories. 

In Table \ref{tab3}, we took as a test the mass of the strange star candidate LMC X-4 ($M=1.29\pm 0.05~M_\odot$) to predict its radius, compactness, surface redshift, central energy density and central pressure for different values of $\sigma$ and $B=60$ MeV/fm$^3$. The LMC X-4 is an object extensively pointed as a strange star candidate due to its structure and characteristics \cite{Weber2012Aug,Deb2018Mar,Carvalho2020Jun,Sharif2018Oct,Majid2020Aug}. 

\section{Conclusions}\label{secVI}

In this work, we have studied strange stars in the background of a GMC theory, particularly, the one described by the gravitational action $f(R,L)=R/2+L+\sigma RL$. The effects of the theory on strange star macroscopic properties are that the stars become larger and more massive, leading to larger compactness and surface redshift. One important feature of the theory is that it recovers GR for energy densities smaller than 250 MeV/fm$^3$ not depending on the value of $\sigma$. 

In addition, the space-time metric outside the stars is described by the exterior Schwarzschild solution, which means that the space-time outside is neither AdS nor de Sitter, as it is in some $f(R)$ models. Also, the GMC theory explored here has, as a property, energy-momentum conservation, which is an advantage compared to various other modified theories of gravity.

In particular, when $\sigma=50$ km$^2$, the GMC theory can achieve  2.6 $M_\odot$, which can describe, e.g., the mass of the pulsars PSR J2215+5135, PSR J1614-2230, PSR J0952-0607 and of the secondary object in the GW190814 event, if those objects are considered to be strange stars. The theory can also describe strange quark star candidates with feasible physical properties, as detailed in Tables \ref{tab2} and \ref{tab3}.

When calculating the mass of the object for the GMC theory described here, we considered that the mass is calculated according to \eqref{gravitmass2}. This means, in particular, that the gravitational mass is obtained by requiring a smooth connection between the interior and exterior metrics, i.e., junction conditions are respected. So, the definition \eqref{gravitmass2} gives the gravitational mass as perceived by a distant observer, which can be compared to observational data. This smooth connection between the interior and exterior metrics is a consequence of the GMC theory because, at the star surface, the pressure vanishes and all the new terms coming from the GMC theory go to zero. Thus, at the star surface, the interior metric must have the form of the Schwarzschild one, so there is not a spherical vacuum layer where the scalar curvature is non-zero \cite{Maurya2019Aug}.

\section*{Acknowledgments}
GAC thanks Coordena\c{c}\~ao de Aperfei\c{c}oamento Pessoal de N\'ivel Superior (CAPES) for financial support under grant PNPD/88887.368365/2019-00. RVL is supported by U.S. Department of Energy (DOE) under grant DE--FG02--08ER41533 and to the LANL Collaborative Research Program by Texas A\&M System National Laboratory Office and Los Alamos National Laboratory. MM acknowledge CAPES, CNPq and  project INCT-FNA Proc. No. 464898/2014-5. PHRSM thanks CAPES for financial support.  The research of DD is funded by the C.V. Raman Postdoctoral Fellowship (Reg. No. R(IA)CVR-PDF/2020/222) from the Department of Physics, Indian Institute of Science

\bibliography{library}

\begin{thebibliography}{84}
\expandafter\ifx\csname natexlab\endcsname\relax\def\natexlab#1{#1}\fi
\expandafter\ifx\csname bibnamefont\endcsname\relax
  \def\bibnamefont#1{#1}\fi
\expandafter\ifx\csname bibfnamefont\endcsname\relax
  \def\bibfnamefont#1{#1}\fi
\expandafter\ifx\csname citenamefont\endcsname\relax
  \def\citenamefont#1{#1}\fi
\expandafter\ifx\csname url\endcsname\relax
  \def\url#1{\texttt{#1}}\fi
\expandafter\ifx\csname urlprefix\endcsname\relax\def\urlprefix{URL }\fi
\providecommand{\bibinfo}[2]{#2}
\providecommand{\eprint}[2][]{\url{#2}}

\bibitem[{\citenamefont{Perlmutter et~al.}(1999)}]{Perlmutter1999}
\bibinfo{author}{\bibfnamefont{S.}~\bibnamefont{Perlmutter}}
  \bibnamefont{et~al.}, \bibinfo{journal}{Astrophys. J.}
  \textbf{\bibinfo{volume}{517}}, \bibinfo{pages}{565} (\bibinfo{year}{1999}),
  ISSN \bibinfo{issn}{0004-637X}.

\bibitem[{\citenamefont{Bennett et~al.}(2003)}]{Bennett2003}
\bibinfo{author}{\bibfnamefont{C.~L.} \bibnamefont{Bennett}}
  \bibnamefont{et~al.}, \bibinfo{journal}{Astrophys. J. Suppl. Ser.}
  \textbf{\bibinfo{volume}{148}}, \bibinfo{pages}{97} (\bibinfo{year}{2003}),
  ISSN \bibinfo{issn}{0067-0049}.

\bibitem[{\citenamefont{Riess et~al.}(1998)}]{Riess1998}
\bibinfo{author}{\bibfnamefont{A.~G.} \bibnamefont{Riess}}
  \bibnamefont{et~al.}, \bibinfo{journal}{Astron. J.}
  \textbf{\bibinfo{volume}{116}}, \bibinfo{pages}{1009} (\bibinfo{year}{1998}),
  ISSN \bibinfo{issn}{0004-6256}.

\bibitem[{\citenamefont{Spergel et~al.}(2003)}]{Spergel2003}
\bibinfo{author}{\bibfnamefont{D.~N.} \bibnamefont{Spergel}}
  \bibnamefont{et~al.}, \bibinfo{journal}{Astrophys. J. Suppl. Ser.}
  \textbf{\bibinfo{volume}{148}}, \bibinfo{pages}{175} (\bibinfo{year}{2003}),
  ISSN \bibinfo{issn}{0067-0049}.

\bibitem[{\citenamefont{Spergel et~al.}(2007)}]{Spergel2007}
\bibinfo{author}{\bibfnamefont{D.~N.} \bibnamefont{Spergel}}
  \bibnamefont{et~al.}, \bibinfo{journal}{Astrophys. J. Suppl. Ser.}
  \textbf{\bibinfo{volume}{170}}, \bibinfo{pages}{377} (\bibinfo{year}{2007}),
  ISSN \bibinfo{issn}{0067-0049}.

\bibitem[{\citenamefont{Ade and others (BICEP2~Collaboration)}(2014)}]{Ade2014}
\bibinfo{author}{\bibfnamefont{P.~A.~R.} \bibnamefont{Ade}} \bibnamefont{and}
  \bibinfo{author}{\bibnamefont{others (BICEP2~Collaboration)}},
  \bibinfo{journal}{Phys. Rev. Lett.} \textbf{\bibinfo{volume}{112}},
  \bibinfo{pages}{241101} (\bibinfo{year}{2014}), ISSN
  \bibinfo{issn}{1079-7114}.

\bibitem[{\citenamefont{Weinberg}(1989)}]{Weinberg1989}
\bibinfo{author}{\bibfnamefont{S.}~\bibnamefont{Weinberg}},
  \bibinfo{journal}{Rev. Mod. Phys.} \textbf{\bibinfo{volume}{61}},
  \bibinfo{pages}{1} (\bibinfo{year}{1989}), ISSN \bibinfo{issn}{1539-0756}.

\bibitem[{\citenamefont{Carroll}(2001)}]{Carroll2001}
\bibinfo{author}{\bibfnamefont{S.~M.} \bibnamefont{Carroll}},
  \bibinfo{journal}{Living Rev. Relativ.} \textbf{\bibinfo{volume}{4}},
  \bibinfo{pages}{1} (\bibinfo{year}{2001}), ISSN \bibinfo{issn}{1433-8351}.

\bibitem[{\citenamefont{Amendola et~al.}(2007)\citenamefont{Amendola, Gannouji,
  Polarski, and Tsujikawa}}]{amendola2007}
\bibinfo{author}{\bibfnamefont{L.}~\bibnamefont{Amendola}},
  \bibinfo{author}{\bibfnamefont{R.}~\bibnamefont{Gannouji}},
  \bibinfo{author}{\bibfnamefont{D.}~\bibnamefont{Polarski}}, \bibnamefont{and}
  \bibinfo{author}{\bibfnamefont{S.}~\bibnamefont{Tsujikawa}},
  \bibinfo{journal}{Phys. Rev. D} \textbf{\bibinfo{volume}{75}},
  \bibinfo{pages}{083504} (\bibinfo{year}{2007}), ISSN
  \bibinfo{issn}{2470-0029}.

\bibitem[{\citenamefont{Hwang and Noh}(2001)}]{hwang/2001}
\bibinfo{author}{\bibfnamefont{J.-c.} \bibnamefont{Hwang}} \bibnamefont{and}
  \bibinfo{author}{\bibfnamefont{H.}~\bibnamefont{Noh}},
  \bibinfo{journal}{Phys. Lett. B} \textbf{\bibinfo{volume}{506}},
  \bibinfo{pages}{13} (\bibinfo{year}{2001}), ISSN \bibinfo{issn}{0370-2693}.

\bibitem[{\citenamefont{Harko and Lobo}(2010)}]{Harko2010}
\bibinfo{author}{\bibfnamefont{T.}~\bibnamefont{Harko}} \bibnamefont{and}
  \bibinfo{author}{\bibfnamefont{F.~S.~N.} \bibnamefont{Lobo}},
  \bibinfo{journal}{Eur. Phys. J. C} \textbf{\bibinfo{volume}{70}},
  \bibinfo{pages}{373} (\bibinfo{year}{2010}), ISSN \bibinfo{issn}{1434-6052}.

\bibitem[{\citenamefont{{Wang} et~al.}(2018)\citenamefont{{Wang}, {Gui}, and
  {Qiu}}}]{wang/2018}
\bibinfo{author}{\bibfnamefont{J.}~\bibnamefont{{Wang}}},
  \bibinfo{author}{\bibfnamefont{R.}~\bibnamefont{{Gui}}}, \bibnamefont{and}
  \bibinfo{author}{\bibfnamefont{W.}~\bibnamefont{{Qiu}}},
  \bibinfo{journal}{Physics of the Dark Universe}
  \textbf{\bibinfo{volume}{19}}, \bibinfo{pages}{60} (\bibinfo{year}{2018}).

\bibitem[{\citenamefont{Nesseris}(2009)}]{nesseris/2009}
\bibinfo{author}{\bibfnamefont{S.}~\bibnamefont{Nesseris}},
  \bibinfo{journal}{Phys. Rev. D} \textbf{\bibinfo{volume}{79}},
  \bibinfo{pages}{044015} (\bibinfo{year}{2009}), ISSN
  \bibinfo{issn}{2470-0029}.

\bibitem[{\citenamefont{Delsate and Steinhoff}(2012)}]{delsate/2012}
\bibinfo{author}{\bibfnamefont{T.}~\bibnamefont{Delsate}} \bibnamefont{and}
  \bibinfo{author}{\bibfnamefont{J.}~\bibnamefont{Steinhoff}},
  \bibinfo{journal}{Phys. Rev. Lett.} \textbf{\bibinfo{volume}{109}},
  \bibinfo{pages}{021101} (\bibinfo{year}{2012}), ISSN
  \bibinfo{issn}{1079-7114}.

\bibitem[{\citenamefont{Azevedo and
  P{\ifmmode\acute{a}\else\'{a}\fi}ramos}(2016)}]{Azevedo2016}
\bibinfo{author}{\bibfnamefont{R.~P.~L.} \bibnamefont{Azevedo}}
  \bibnamefont{and}
  \bibinfo{author}{\bibfnamefont{J.}~\bibnamefont{P{\ifmmode\acute{a}\else\'{a}\fi}ramos}},
  \bibinfo{journal}{Phys. Rev. D} \textbf{\bibinfo{volume}{94}},
  \bibinfo{pages}{064036} (\bibinfo{year}{2016}), ISSN
  \bibinfo{issn}{2470-0029}.

\bibitem[{\citenamefont{Avelino and Azevedo}(2018)}]{Avelino2018}
\bibinfo{author}{\bibfnamefont{P.~P.} \bibnamefont{Avelino}} \bibnamefont{and}
  \bibinfo{author}{\bibfnamefont{R.~P.~L.} \bibnamefont{Azevedo}},
  \bibinfo{journal}{Phys. Rev. D} \textbf{\bibinfo{volume}{97}},
  \bibinfo{pages}{064018} (\bibinfo{year}{2018}), ISSN
  \bibinfo{issn}{2470-0029}.

\bibitem[{\citenamefont{Wang and Liao}(2012)}]{Wang2012}
\bibinfo{author}{\bibfnamefont{J.}~\bibnamefont{Wang}} \bibnamefont{and}
  \bibinfo{author}{\bibfnamefont{K.}~\bibnamefont{Liao}},
  \bibinfo{journal}{Classical Quantum Gravity} \textbf{\bibinfo{volume}{29}},
  \bibinfo{pages}{215016} (\bibinfo{year}{2012}), ISSN
  \bibinfo{issn}{0264-9381}.

\bibitem[{\citenamefont{Wu et~al.}(2014)\citenamefont{Wu, Zhao, Jin, Lin, Lu,
  and Zhang}}]{Wu2014}
\bibinfo{author}{\bibfnamefont{Y.-B.} \bibnamefont{Wu}},
  \bibinfo{author}{\bibfnamefont{Y.-Y.} \bibnamefont{Zhao}},
  \bibinfo{author}{\bibfnamefont{Y.-Y.} \bibnamefont{Jin}},
  \bibinfo{author}{\bibfnamefont{L.-L.} \bibnamefont{Lin}},
  \bibinfo{author}{\bibfnamefont{J.-B.} \bibnamefont{Lu}}, \bibnamefont{and}
  \bibinfo{author}{\bibfnamefont{X.}~\bibnamefont{Zhang}},
  \bibinfo{journal}{Mod. Phys. Lett. A} \textbf{\bibinfo{volume}{29}}
  (\bibinfo{year}{2014}).

\bibitem[{\citenamefont{Harko et~al.}(2015)\citenamefont{Harko, Lobo, Mimoso,
  and Pav{\ifmmode\acute{o}\else\'{o}\fi}n}}]{Harko2015}
\bibinfo{author}{\bibfnamefont{T.}~\bibnamefont{Harko}},
  \bibinfo{author}{\bibfnamefont{F.~S.~N.} \bibnamefont{Lobo}},
  \bibinfo{author}{\bibfnamefont{J.~P.} \bibnamefont{Mimoso}},
  \bibnamefont{and}
  \bibinfo{author}{\bibfnamefont{D.}~\bibnamefont{Pav{\ifmmode\acute{o}\else\'{o}\fi}n}},
  \bibinfo{journal}{Eur. Phys. J. C} \textbf{\bibinfo{volume}{75}},
  \bibinfo{pages}{386} (\bibinfo{year}{2015}), ISSN \bibinfo{issn}{1434-6052}.

\bibitem[{\citenamefont{Harko}(2014)}]{Harko2014D}
\bibinfo{author}{\bibfnamefont{T.}~\bibnamefont{Harko}},
  \bibinfo{journal}{Phys. Rev. D} \textbf{\bibinfo{volume}{90}},
  \bibinfo{pages}{044067} (\bibinfo{year}{2014}), ISSN
  \bibinfo{issn}{2470-0029}.

\bibitem[{\citenamefont{Pourhassan and Rudra}(2020)}]{pourhassan/2020}
\bibinfo{author}{\bibfnamefont{B.}~\bibnamefont{Pourhassan}} \bibnamefont{and}
  \bibinfo{author}{\bibfnamefont{P.}~\bibnamefont{Rudra}},
  \bibinfo{journal}{Phys. Rev. D} \textbf{\bibinfo{volume}{101}},
  \bibinfo{pages}{084057} (\bibinfo{year}{2020}), ISSN
  \bibinfo{issn}{2470-0029}.

\bibitem[{\citenamefont{Avelino and Sousa}(2018)}]{Avelino2018D}
\bibinfo{author}{\bibfnamefont{P.~P.} \bibnamefont{Avelino}} \bibnamefont{and}
  \bibinfo{author}{\bibfnamefont{L.}~\bibnamefont{Sousa}},
  \bibinfo{journal}{Phys. Rev. D} \textbf{\bibinfo{volume}{97}},
  \bibinfo{pages}{064019} (\bibinfo{year}{2018}), ISSN
  \bibinfo{issn}{2470-0029}.

\bibitem[{\citenamefont{Harko et~al.}(2011)\citenamefont{Harko, Lobo, Nojiri,
  and Odintsov}}]{Harko2011}
\bibinfo{author}{\bibfnamefont{T.}~\bibnamefont{Harko}},
  \bibinfo{author}{\bibfnamefont{F.~S.~N.} \bibnamefont{Lobo}},
  \bibinfo{author}{\bibfnamefont{S.}~\bibnamefont{Nojiri}}, \bibnamefont{and}
  \bibinfo{author}{\bibfnamefont{S.~D.} \bibnamefont{Odintsov}},
  \bibinfo{journal}{Phys. Rev. D} \textbf{\bibinfo{volume}{84}},
  \bibinfo{pages}{024020} (\bibinfo{year}{2011}), ISSN
  \bibinfo{issn}{2470-0029}.

\bibitem[{\citenamefont{Harko et~al.}(2013)\citenamefont{Harko, Lobo, and
  Minazzoli}}]{Harko2013}
\bibinfo{author}{\bibfnamefont{T.}~\bibnamefont{Harko}},
  \bibinfo{author}{\bibfnamefont{F.~S.~N.} \bibnamefont{Lobo}},
  \bibnamefont{and}
  \bibinfo{author}{\bibfnamefont{O.}~\bibnamefont{Minazzoli}},
  \bibinfo{journal}{Phys. Rev. D} \textbf{\bibinfo{volume}{87}},
  \bibinfo{pages}{047501} (\bibinfo{year}{2013}), ISSN
  \bibinfo{issn}{2470-0029}.

\bibitem[{\citenamefont{Harko}(2008)}]{Harko2008}
\bibinfo{author}{\bibfnamefont{T.}~\bibnamefont{Harko}},
  \bibinfo{journal}{Phys. Lett. B} \textbf{\bibinfo{volume}{669}},
  \bibinfo{pages}{376} (\bibinfo{year}{2008}), ISSN \bibinfo{issn}{0370-2693}.

\bibitem[{\citenamefont{Moraes and Sahoo}(2017)}]{Moraes2017}
\bibinfo{author}{\bibfnamefont{P.~H. R.~S.} \bibnamefont{Moraes}}
  \bibnamefont{and} \bibinfo{author}{\bibfnamefont{P.~K.} \bibnamefont{Sahoo}},
  \bibinfo{journal}{Eur. Phys. J. C} \textbf{\bibinfo{volume}{77}},
  \bibinfo{pages}{480} (\bibinfo{year}{2017}), ISSN \bibinfo{issn}{1434-6052}.

\bibitem[{\citenamefont{Harko}(2010)}]{Harko2010D}
\bibinfo{author}{\bibfnamefont{T.}~\bibnamefont{Harko}},
  \bibinfo{journal}{Phys. Rev. D} \textbf{\bibinfo{volume}{81}},
  \bibinfo{pages}{044021} (\bibinfo{year}{2010}), ISSN
  \bibinfo{issn}{2470-0029}.

\bibitem[{\citenamefont{Copeland et~al.}(2006)\citenamefont{Copeland, Sami, and
  Tsujikawa}}]{Copeland2006}
\bibinfo{author}{\bibfnamefont{E.~J.} \bibnamefont{Copeland}},
  \bibinfo{author}{\bibfnamefont{M.}~\bibnamefont{Sami}}, \bibnamefont{and}
  \bibinfo{author}{\bibfnamefont{S.}~\bibnamefont{Tsujikawa}},
  \bibinfo{journal}{Int. J. Mod. Phys. D} \textbf{\bibinfo{volume}{15}},
  \bibinfo{pages}{1753} (\bibinfo{year}{2006}), ISSN \bibinfo{issn}{0218-2718}.

\bibitem[{\citenamefont{Frieman et~al.}(2008)\citenamefont{Frieman, Turner, and
  Huterer}}]{Frieman2008}
\bibinfo{author}{\bibfnamefont{J.~A.} \bibnamefont{Frieman}},
  \bibinfo{author}{\bibfnamefont{M.~S.} \bibnamefont{Turner}},
  \bibnamefont{and} \bibinfo{author}{\bibfnamefont{D.}~\bibnamefont{Huterer}},
  \bibinfo{journal}{Annu. Rev. Astron. Astrophys.}
  \textbf{\bibinfo{volume}{46}}, \bibinfo{pages}{385} (\bibinfo{year}{2008}),
  ISSN \bibinfo{issn}{0066-4146}.

\bibitem[{\citenamefont{{Navarro} et~al.}(1996)\citenamefont{{Navarro},
  {Frenk}, and {White}}}]{navarro/1996}
\bibinfo{author}{\bibfnamefont{J.~F.} \bibnamefont{{Navarro}}},
  \bibinfo{author}{\bibfnamefont{C.~S.} \bibnamefont{{Frenk}}},
  \bibnamefont{and} \bibinfo{author}{\bibfnamefont{S.~D.~M.}
  \bibnamefont{{White}}}, \bibinfo{journal}{Astrophys. J.}
  \textbf{\bibinfo{volume}{462}}, \bibinfo{pages}{563} (\bibinfo{year}{1996}),
  \eprint{astro-ph/9508025}.

\bibitem[{\citenamefont{{Moore} et~al.}(1999)\citenamefont{{Moore}, {Ghigna},
  {Governato}, {Lake}, {Quinn}, {Stadel}, and {Tozzi}}}]{moore/1999}
\bibinfo{author}{\bibfnamefont{B.}~\bibnamefont{{Moore}}},
  \bibinfo{author}{\bibfnamefont{S.}~\bibnamefont{{Ghigna}}},
  \bibinfo{author}{\bibfnamefont{F.}~\bibnamefont{{Governato}}},
  \bibinfo{author}{\bibfnamefont{G.}~\bibnamefont{{Lake}}},
  \bibinfo{author}{\bibfnamefont{T.}~\bibnamefont{{Quinn}}},
  \bibinfo{author}{\bibfnamefont{J.}~\bibnamefont{{Stadel}}}, \bibnamefont{and}
  \bibinfo{author}{\bibfnamefont{P.}~\bibnamefont{{Tozzi}}},
  \bibinfo{journal}{Astrophys. J. Lett.} \textbf{\bibinfo{volume}{524}},
  \bibinfo{pages}{L19} (\bibinfo{year}{1999}), \eprint{astro-ph/9907411}.

\bibitem[{\citenamefont{Bertone et~al.}(2005)\citenamefont{Bertone, Hooper, and
  Silk}}]{bertone/2005}
\bibinfo{author}{\bibfnamefont{G.}~\bibnamefont{Bertone}},
  \bibinfo{author}{\bibfnamefont{D.}~\bibnamefont{Hooper}}, \bibnamefont{and}
  \bibinfo{author}{\bibfnamefont{J.}~\bibnamefont{Silk}},
  \bibinfo{journal}{Phys. Rep.} \textbf{\bibinfo{volume}{405}},
  \bibinfo{pages}{279} (\bibinfo{year}{2005}), ISSN \bibinfo{issn}{0370-1573}.

\bibitem[{\citenamefont{Harko and Lobo}(2014{\natexlab{a}})}]{Harko2014E}
\bibinfo{author}{\bibfnamefont{T.}~\bibnamefont{Harko}} \bibnamefont{and}
  \bibinfo{author}{\bibfnamefont{F.~S.~N.} \bibnamefont{Lobo}},
  \bibinfo{journal}{Galaxies} \textbf{\bibinfo{volume}{2}},
  \bibinfo{pages}{410} (\bibinfo{year}{2014}{\natexlab{a}}), ISSN
  \bibinfo{issn}{2075-4434}.

\bibitem[{\citenamefont{Carvalho
  et~al.}(2020{\natexlab{a}})\citenamefont{Carvalho, Moraes, dos Santos,
  Gon{\ifmmode\mbox{\c{c}}\else\c{c}\fi}alves, and Malheiro}}]{Carvalho2020May}
\bibinfo{author}{\bibfnamefont{G.~A.} \bibnamefont{Carvalho}},
  \bibinfo{author}{\bibfnamefont{P.~H. R.~S.} \bibnamefont{Moraes}},
  \bibinfo{author}{\bibfnamefont{S.~I.} \bibnamefont{dos Santos}},
  \bibinfo{author}{\bibfnamefont{B.~S.}
  \bibnamefont{Gon{\ifmmode\mbox{\c{c}}\else\c{c}\fi}alves}}, \bibnamefont{and}
  \bibinfo{author}{\bibfnamefont{M.}~\bibnamefont{Malheiro}},
  \bibinfo{journal}{Eur. Phys. J. C} \textbf{\bibinfo{volume}{80}},
  \bibinfo{pages}{483} (\bibinfo{year}{2020}{\natexlab{a}}), ISSN
  \bibinfo{issn}{1434-6052}.

\bibitem[{\citenamefont{Lobato et~al.}(2021)\citenamefont{Lobato, Carvalho, and
  Bertulani}}]{Lobato2021Nov}
\bibinfo{author}{\bibfnamefont{R.~V.} \bibnamefont{Lobato}},
  \bibinfo{author}{\bibfnamefont{G.~A.} \bibnamefont{Carvalho}},
  \bibnamefont{and} \bibinfo{author}{\bibfnamefont{C.~A.}
  \bibnamefont{Bertulani}}, \bibinfo{journal}{Eur. Phys. J. C}
  \textbf{\bibinfo{volume}{81}}, \bibinfo{pages}{1} (\bibinfo{year}{2021}),
  ISSN \bibinfo{issn}{1434-6052}.

\bibitem[{\citenamefont{Riley et~al.}(2019)\citenamefont{Riley, Watts,
  Bogdanov, Ray, Ludlam, Guillot, Arzoumanian, Baker, Bilous, Chakrabarty
  et~al.}}]{Riley2019Dec}
\bibinfo{author}{\bibfnamefont{T.~E.} \bibnamefont{Riley}},
  \bibinfo{author}{\bibfnamefont{A.~L.} \bibnamefont{Watts}},
  \bibinfo{author}{\bibfnamefont{S.}~\bibnamefont{Bogdanov}},
  \bibinfo{author}{\bibfnamefont{P.~S.} \bibnamefont{Ray}},
  \bibinfo{author}{\bibfnamefont{R.~M.} \bibnamefont{Ludlam}},
  \bibinfo{author}{\bibfnamefont{S.}~\bibnamefont{Guillot}},
  \bibinfo{author}{\bibfnamefont{Z.}~\bibnamefont{Arzoumanian}},
  \bibinfo{author}{\bibfnamefont{C.~L.} \bibnamefont{Baker}},
  \bibinfo{author}{\bibfnamefont{A.~V.} \bibnamefont{Bilous}},
  \bibinfo{author}{\bibfnamefont{D.}~\bibnamefont{Chakrabarty}},
  \bibnamefont{et~al.}, \bibinfo{journal}{Astrophys. J. Lett.}
  \textbf{\bibinfo{volume}{887}}, \bibinfo{pages}{L21} (\bibinfo{year}{2019}),
  ISSN \bibinfo{issn}{2041-8213}.

\bibitem[{\citenamefont{Miller et~al.}(2021)\citenamefont{Miller, Lamb,
  Dittmann, Bogdanov, Arzoumanian, Gendreau, Guillot, Ho, Lattimer, Loewenstein
  et~al.}}]{Miller2021Sep}
\bibinfo{author}{\bibfnamefont{M.~C.} \bibnamefont{Miller}},
  \bibinfo{author}{\bibfnamefont{F.~K.} \bibnamefont{Lamb}},
  \bibinfo{author}{\bibfnamefont{A.~J.} \bibnamefont{Dittmann}},
  \bibinfo{author}{\bibfnamefont{S.}~\bibnamefont{Bogdanov}},
  \bibinfo{author}{\bibfnamefont{Z.}~\bibnamefont{Arzoumanian}},
  \bibinfo{author}{\bibfnamefont{K.~C.} \bibnamefont{Gendreau}},
  \bibinfo{author}{\bibfnamefont{S.}~\bibnamefont{Guillot}},
  \bibinfo{author}{\bibfnamefont{W.~C.~G.} \bibnamefont{Ho}},
  \bibinfo{author}{\bibfnamefont{J.~M.} \bibnamefont{Lattimer}},
  \bibinfo{author}{\bibfnamefont{M.}~\bibnamefont{Loewenstein}},
  \bibnamefont{et~al.}, \bibinfo{journal}{Astrophys. J. Lett.}
  \textbf{\bibinfo{volume}{918}}, \bibinfo{pages}{L28} (\bibinfo{year}{2021}),
  ISSN \bibinfo{issn}{2041-8205}.

\bibitem[{\citenamefont{Riley et~al.}(2021)\citenamefont{Riley, Watts, Ray,
  Bogdanov, Guillot, Morsink, Bilous, Arzoumanian, Choudhury, Deneva
  et~al.}}]{Riley2021Sep}
\bibinfo{author}{\bibfnamefont{T.~E.} \bibnamefont{Riley}},
  \bibinfo{author}{\bibfnamefont{A.~L.} \bibnamefont{Watts}},
  \bibinfo{author}{\bibfnamefont{P.~S.} \bibnamefont{Ray}},
  \bibinfo{author}{\bibfnamefont{S.}~\bibnamefont{Bogdanov}},
  \bibinfo{author}{\bibfnamefont{S.}~\bibnamefont{Guillot}},
  \bibinfo{author}{\bibfnamefont{S.~M.} \bibnamefont{Morsink}},
  \bibinfo{author}{\bibfnamefont{A.~V.} \bibnamefont{Bilous}},
  \bibinfo{author}{\bibfnamefont{Z.}~\bibnamefont{Arzoumanian}},
  \bibinfo{author}{\bibfnamefont{D.}~\bibnamefont{Choudhury}},
  \bibinfo{author}{\bibfnamefont{J.~S.} \bibnamefont{Deneva}},
  \bibnamefont{et~al.}, \bibinfo{journal}{Astrophys. J. Lett.}
  \textbf{\bibinfo{volume}{918}}, \bibinfo{pages}{L27} (\bibinfo{year}{2021}),
  ISSN \bibinfo{issn}{2041-8205}.

\bibitem[{\citenamefont{Abbott et~al.}(2017)}]{abbott/2017}
\bibinfo{author}{\bibfnamefont{B.~P.} \bibnamefont{Abbott}}
  \bibnamefont{et~al.}, \bibinfo{journal}{Phys. Rev. Lett.}
  \textbf{\bibinfo{volume}{119}}, \bibinfo{pages}{161101}
  (\bibinfo{year}{2017}), ISSN \bibinfo{issn}{1079-7114}.

\bibitem[{\citenamefont{Romani et~al.}(2022)\citenamefont{Romani, Kandel,
  Filippenko, Brink, and Zheng}}]{Romani2022Jul}
\bibinfo{author}{\bibfnamefont{R.~W.} \bibnamefont{Romani}},
  \bibinfo{author}{\bibfnamefont{D.}~\bibnamefont{Kandel}},
  \bibinfo{author}{\bibfnamefont{A.~V.} \bibnamefont{Filippenko}},
  \bibinfo{author}{\bibfnamefont{T.~G.} \bibnamefont{Brink}}, \bibnamefont{and}
  \bibinfo{author}{\bibfnamefont{W.}~\bibnamefont{Zheng}},
  \bibinfo{journal}{Astrophys. J. Lett.} \textbf{\bibinfo{volume}{934}},
  \bibinfo{pages}{L17} (\bibinfo{year}{2022}), ISSN \bibinfo{issn}{2041-8205}.

\bibitem[{\citenamefont{Bassa et~al.}(2017)\citenamefont{Bassa, Pleunis,
  Hessels, Ferrara, Breton, Gusinskaia, Kondratiev, Sanidas, Nieder, Clark
  et~al.}}]{Bassa2017Sep}
\bibinfo{author}{\bibfnamefont{C.~G.} \bibnamefont{Bassa}},
  \bibinfo{author}{\bibfnamefont{Z.}~\bibnamefont{Pleunis}},
  \bibinfo{author}{\bibfnamefont{J.~W.~T.} \bibnamefont{Hessels}},
  \bibinfo{author}{\bibfnamefont{E.~C.} \bibnamefont{Ferrara}},
  \bibinfo{author}{\bibfnamefont{R.~P.} \bibnamefont{Breton}},
  \bibinfo{author}{\bibfnamefont{N.~V.} \bibnamefont{Gusinskaia}},
  \bibinfo{author}{\bibfnamefont{V.~I.} \bibnamefont{Kondratiev}},
  \bibinfo{author}{\bibfnamefont{S.}~\bibnamefont{Sanidas}},
  \bibinfo{author}{\bibfnamefont{L.}~\bibnamefont{Nieder}},
  \bibinfo{author}{\bibfnamefont{C.~J.} \bibnamefont{Clark}},
  \bibnamefont{et~al.}, \bibinfo{journal}{Astrophys. J. Lett.}
  \textbf{\bibinfo{volume}{846}}, \bibinfo{pages}{L20} (\bibinfo{year}{2017}),
  ISSN \bibinfo{issn}{2041-8213}.

\bibitem[{\citenamefont{Farhi and Jaffe}(1984{\natexlab{a}})}]{Farhi1984}
\bibinfo{author}{\bibfnamefont{E.}~\bibnamefont{Farhi}} \bibnamefont{and}
  \bibinfo{author}{\bibfnamefont{R.~L.} \bibnamefont{Jaffe}},
  \bibinfo{journal}{Phys. Rev. D} \textbf{\bibinfo{volume}{30}},
  \bibinfo{pages}{2379} (\bibinfo{year}{1984}{\natexlab{a}}), ISSN
  \bibinfo{issn}{2470-0029}.

\bibitem[{\citenamefont{Alcock and Olinto}(1988)}]{Alcock1988}
\bibinfo{author}{\bibfnamefont{C.}~\bibnamefont{Alcock}} \bibnamefont{and}
  \bibinfo{author}{\bibfnamefont{A.}~\bibnamefont{Olinto}},
  \bibinfo{journal}{Annu. Rev. Nucl. Part. Sci.} \textbf{\bibinfo{volume}{38}},
  \bibinfo{pages}{161} (\bibinfo{year}{1988}), ISSN \bibinfo{issn}{0163-8998}.

\bibitem[{\citenamefont{Bodmer}(1971)}]{Bodmer1971}
\bibinfo{author}{\bibfnamefont{A.~R.} \bibnamefont{Bodmer}},
  \bibinfo{journal}{Phys. Rev. D} \textbf{\bibinfo{volume}{4}},
  \bibinfo{pages}{1601} (\bibinfo{year}{1971}), ISSN \bibinfo{issn}{2470-0029}.

\bibitem[{\citenamefont{Witten}(1984)}]{Witten1984}
\bibinfo{author}{\bibfnamefont{E.}~\bibnamefont{Witten}},
  \bibinfo{journal}{Phys. Rev. D} \textbf{\bibinfo{volume}{30}},
  \bibinfo{pages}{272} (\bibinfo{year}{1984}), ISSN \bibinfo{issn}{2470-0029}.

\bibitem[{\citenamefont{Malheiro et~al.}(2003)\citenamefont{Malheiro, Fiolhais,
  and Taurines}}]{Malheiro2003Apr}
\bibinfo{author}{\bibfnamefont{M.}~\bibnamefont{Malheiro}},
  \bibinfo{author}{\bibfnamefont{M.}~\bibnamefont{Fiolhais}}, \bibnamefont{and}
  \bibinfo{author}{\bibfnamefont{A.~R.} \bibnamefont{Taurines}},
  \bibinfo{journal}{J. Phys. G: Nucl. Part. Phys.}
  \textbf{\bibinfo{volume}{29}}, \bibinfo{pages}{1045} (\bibinfo{year}{2003}),
  ISSN \bibinfo{issn}{0954-3899}.

\bibitem[{\citenamefont{Moraes and Miranda}(2014)}]{moraes/2014}
\bibinfo{author}{\bibfnamefont{P.~H. R.~S.} \bibnamefont{Moraes}}
  \bibnamefont{and} \bibinfo{author}{\bibfnamefont{O.~D.}
  \bibnamefont{Miranda}}, \bibinfo{journal}{Mon. Not. R. Astron. Soc. Lett.}
  \textbf{\bibinfo{volume}{445}}, \bibinfo{pages}{L11} (\bibinfo{year}{2014}),
  ISSN \bibinfo{issn}{1745-3925}.

\bibitem[{\citenamefont{Weber}(2005)}]{Weber2005}
\bibinfo{author}{\bibfnamefont{F.}~\bibnamefont{Weber}},
  \bibinfo{journal}{Prog. Part. Nucl. Phys.} \textbf{\bibinfo{volume}{54}},
  \bibinfo{pages}{193} (\bibinfo{year}{2005}), ISSN \bibinfo{issn}{0146-6410}.

\bibitem[{\citenamefont{Panotopoulos and
  Rinc{\ifmmode\acute{o}\else\'{o}\fi}n}(2019)}]{Panotopoulos2019}
\bibinfo{author}{\bibfnamefont{G.}~\bibnamefont{Panotopoulos}}
  \bibnamefont{and}
  \bibinfo{author}{\bibfnamefont{{\ifmmode\acute{A}\else\'{A}\fi}.}~\bibnamefont{Rinc{\ifmmode\acute{o}\else\'{o}\fi}n}},
  \bibinfo{journal}{Eur. Phys. J. C} \textbf{\bibinfo{volume}{79}},
  \bibinfo{pages}{524} (\bibinfo{year}{2019}), ISSN \bibinfo{issn}{1434-6052}.

\bibitem[{\citenamefont{Zhou et~al.}(2018)\citenamefont{Zhou, Zhou, and
  Li}}]{zhou/2018}
\bibinfo{author}{\bibfnamefont{E.-P.} \bibnamefont{Zhou}},
  \bibinfo{author}{\bibfnamefont{X.}~\bibnamefont{Zhou}}, \bibnamefont{and}
  \bibinfo{author}{\bibfnamefont{A.}~\bibnamefont{Li}}, \bibinfo{journal}{Phys.
  Rev. D} \textbf{\bibinfo{volume}{97}}, \bibinfo{pages}{083015}
  (\bibinfo{year}{2018}), ISSN \bibinfo{issn}{2470-0029}.

\bibitem[{\citenamefont{Gomes et~al.}(2019)\citenamefont{Gomes, Char, and
  Schramm}}]{gomes/2019}
\bibinfo{author}{\bibfnamefont{R.~O.} \bibnamefont{Gomes}},
  \bibinfo{author}{\bibfnamefont{P.}~\bibnamefont{Char}}, \bibnamefont{and}
  \bibinfo{author}{\bibfnamefont{S.}~\bibnamefont{Schramm}},
  \bibinfo{journal}{Astrophys. J.} \textbf{\bibinfo{volume}{877}},
  \bibinfo{pages}{139} (\bibinfo{year}{2019}), ISSN \bibinfo{issn}{1538-4357}.

\bibitem[{\citenamefont{Garcia and Lobo}(2010)}]{Garcia2010}
\bibinfo{author}{\bibfnamefont{N.~M.} \bibnamefont{Garcia}} \bibnamefont{and}
  \bibinfo{author}{\bibfnamefont{F.~S.~N.} \bibnamefont{Lobo}},
  \bibinfo{journal}{Phys. Rev. D} \textbf{\bibinfo{volume}{82}},
  \bibinfo{pages}{104018} (\bibinfo{year}{2010}), ISSN
  \bibinfo{issn}{2470-0029}.

\bibitem[{\citenamefont{Garcia and Lobo}(2011)}]{Garcia2011}
\bibinfo{author}{\bibfnamefont{N.~M.} \bibnamefont{Garcia}} \bibnamefont{and}
  \bibinfo{author}{\bibfnamefont{F.~S.~N.} \bibnamefont{Lobo}},
  \bibinfo{journal}{Classical Quantum Gravity} \textbf{\bibinfo{volume}{28}},
  \bibinfo{pages}{085018} (\bibinfo{year}{2011}), ISSN
  \bibinfo{issn}{0264-9381}.

\bibitem[{\citenamefont{Harko and Lobo}(2014{\natexlab{b}})}]{Harko2014}
\bibinfo{author}{\bibfnamefont{T.}~\bibnamefont{Harko}} \bibnamefont{and}
  \bibinfo{author}{\bibfnamefont{F.~S.~N.} \bibnamefont{Lobo}},
  \bibinfo{journal}{Galaxies} \textbf{\bibinfo{volume}{2}},
  \bibinfo{pages}{410} (\bibinfo{year}{2014}{\natexlab{b}}), ISSN
  \bibinfo{issn}{2075-4434}.

\bibitem[{\citenamefont{Lobato et~al.}(2022)\citenamefont{Lobato, Carvalho,
  Kelkar, and Nowakowski}}]{Lobato2022Jun}
\bibinfo{author}{\bibfnamefont{R.~V.} \bibnamefont{Lobato}},
  \bibinfo{author}{\bibfnamefont{G.~A.} \bibnamefont{Carvalho}},
  \bibinfo{author}{\bibfnamefont{N.~G.} \bibnamefont{Kelkar}},
  \bibnamefont{and}
  \bibinfo{author}{\bibfnamefont{M.}~\bibnamefont{Nowakowski}},
  \bibinfo{journal}{Eur. Phys. J. C} \textbf{\bibinfo{volume}{82}},
  \bibinfo{pages}{1} (\bibinfo{year}{2022}), ISSN \bibinfo{issn}{1434-6052}.

\bibitem[{\citenamefont{Farhi and Jaffe}(1984{\natexlab{b}})}]{Farhi1984Dec}
\bibinfo{author}{\bibfnamefont{E.}~\bibnamefont{Farhi}} \bibnamefont{and}
  \bibinfo{author}{\bibfnamefont{R.~L.} \bibnamefont{Jaffe}},
  \bibinfo{journal}{Phys. Rev. D} \textbf{\bibinfo{volume}{30}},
  \bibinfo{pages}{2379} (\bibinfo{year}{1984}{\natexlab{b}}), ISSN
  \bibinfo{issn}{2470-0029}.

\bibitem[{\citenamefont{{Moraes} et~al.}(2016)\citenamefont{{Moraes},
  {Arba{\~n}il}, and {Malheiro}}}]{moraes/2016}
\bibinfo{author}{\bibfnamefont{P.~H.~R.~S.} \bibnamefont{{Moraes}}},
  \bibinfo{author}{\bibfnamefont{J.~D.~V.} \bibnamefont{{Arba{\~n}il}}},
  \bibnamefont{and}
  \bibinfo{author}{\bibfnamefont{M.}~\bibnamefont{{Malheiro}}},
  \bibinfo{journal}{J. Cosmol. Astropart. Phys.}
  \textbf{\bibinfo{volume}{2016}}, \bibinfo{eid}{005} (\bibinfo{year}{2016}),
  \eprint{1511.06282}.

\bibitem[{\citenamefont{Jaffe and Low}(1979)}]{Jaffe1979Apr}
\bibinfo{author}{\bibfnamefont{R.~L.} \bibnamefont{Jaffe}} \bibnamefont{and}
  \bibinfo{author}{\bibfnamefont{F.~E.} \bibnamefont{Low}},
  \bibinfo{journal}{Phys. Rev. D} \textbf{\bibinfo{volume}{19}},
  \bibinfo{pages}{2105} (\bibinfo{year}{1979}), ISSN \bibinfo{issn}{2470-0029}.

\bibitem[{\citenamefont{{Linares} et~al.}(2018)\citenamefont{{Linares},
  {Shahbaz}, and {Casares}}}]{linares/2018}
\bibinfo{author}{\bibfnamefont{M.}~\bibnamefont{{Linares}}},
  \bibinfo{author}{\bibfnamefont{T.}~\bibnamefont{{Shahbaz}}},
  \bibnamefont{and}
  \bibinfo{author}{\bibfnamefont{J.}~\bibnamefont{{Casares}}},
  \bibinfo{journal}{\apj} \textbf{\bibinfo{volume}{859}}, \bibinfo{eid}{54}
  (\bibinfo{year}{2018}), \eprint{1805.08799}.

\bibitem[{\citenamefont{Demorest et~al.}(2010)\citenamefont{Demorest, Pennucci,
  Ransom, Roberts, and Hessels}}]{Demorest2010Oct}
\bibinfo{author}{\bibfnamefont{P.~B.} \bibnamefont{Demorest}},
  \bibinfo{author}{\bibfnamefont{T.}~\bibnamefont{Pennucci}},
  \bibinfo{author}{\bibfnamefont{S.~M.} \bibnamefont{Ransom}},
  \bibinfo{author}{\bibfnamefont{M.~S.~E.} \bibnamefont{Roberts}},
  \bibnamefont{and} \bibinfo{author}{\bibfnamefont{J.~W.~T.}
  \bibnamefont{Hessels}}, \bibinfo{journal}{Nature}
  \textbf{\bibinfo{volume}{467}}, \bibinfo{pages}{1081} (\bibinfo{year}{2010}),
  ISSN \bibinfo{issn}{1476-4687}.

\bibitem[{\citenamefont{Rawls et~al.}(2011)\citenamefont{Rawls, Orosz,
  McClintock, Torres, Bailyn, and Buxton}}]{Rawls2011Feb}
\bibinfo{author}{\bibfnamefont{M.~L.} \bibnamefont{Rawls}},
  \bibinfo{author}{\bibfnamefont{J.~A.} \bibnamefont{Orosz}},
  \bibinfo{author}{\bibfnamefont{J.~E.} \bibnamefont{McClintock}},
  \bibinfo{author}{\bibfnamefont{M.~A.~P.} \bibnamefont{Torres}},
  \bibinfo{author}{\bibfnamefont{C.~D.} \bibnamefont{Bailyn}},
  \bibnamefont{and} \bibinfo{author}{\bibfnamefont{M.~M.}
  \bibnamefont{Buxton}}, \bibinfo{journal}{Astrophys. J.}
  \textbf{\bibinfo{volume}{730}}, \bibinfo{pages}{25} (\bibinfo{year}{2011}),
  ISSN \bibinfo{issn}{0004-637X}.

\bibitem[{\citenamefont{G{\ifmmode\ddot{u}\else\"{u}\fi}ver
  et~al.}(2010{\natexlab{a}})\citenamefont{G{\ifmmode\ddot{u}\else\"{u}\fi}ver,
  {\ifmmode\ddot{O}\else\"{O}\fi}zel, Cabrera-Lavers, and
  Wroblewski}}]{Guver2010Mar}
\bibinfo{author}{\bibfnamefont{T.}~\bibnamefont{G{\ifmmode\ddot{u}\else\"{u}\fi}ver}},
  \bibinfo{author}{\bibfnamefont{F.}~\bibnamefont{{\ifmmode\ddot{O}\else\"{O}\fi}zel}},
  \bibinfo{author}{\bibfnamefont{A.}~\bibnamefont{Cabrera-Lavers}},
  \bibnamefont{and}
  \bibinfo{author}{\bibfnamefont{P.}~\bibnamefont{Wroblewski}},
  \bibinfo{journal}{Astrophys. J.} \textbf{\bibinfo{volume}{712}},
  \bibinfo{pages}{964} (\bibinfo{year}{2010}{\natexlab{a}}), ISSN
  \bibinfo{issn}{0004-637X}.

\bibitem[{\citenamefont{Freire et~al.}(2011)\citenamefont{Freire, Bassa, Wex,
  Stairs, Champion, Ransom, Lazarus, Kaspi, Hessels, Kramer
  et~al.}}]{Freire2011Apr}
\bibinfo{author}{\bibfnamefont{P.~C.~C.} \bibnamefont{Freire}},
  \bibinfo{author}{\bibfnamefont{C.~G.} \bibnamefont{Bassa}},
  \bibinfo{author}{\bibfnamefont{N.}~\bibnamefont{Wex}},
  \bibinfo{author}{\bibfnamefont{I.~H.} \bibnamefont{Stairs}},
  \bibinfo{author}{\bibfnamefont{D.~J.} \bibnamefont{Champion}},
  \bibinfo{author}{\bibfnamefont{S.~M.} \bibnamefont{Ransom}},
  \bibinfo{author}{\bibfnamefont{P.}~\bibnamefont{Lazarus}},
  \bibinfo{author}{\bibfnamefont{V.~M.} \bibnamefont{Kaspi}},
  \bibinfo{author}{\bibfnamefont{J.~W.~T.} \bibnamefont{Hessels}},
  \bibinfo{author}{\bibfnamefont{M.}~\bibnamefont{Kramer}},
  \bibnamefont{et~al.}, \bibinfo{journal}{Mon. Not. R. Astron. Soc.}
  \textbf{\bibinfo{volume}{412}}, \bibinfo{pages}{2763} (\bibinfo{year}{2011}),
  ISSN \bibinfo{issn}{0035-8711}.

\bibitem[{\citenamefont{G{\ifmmode\ddot{u}\else\"{u}\fi}ver
  et~al.}(2010{\natexlab{b}})\citenamefont{G{\ifmmode\ddot{u}\else\"{u}\fi}ver,
  Wroblewski, Camarota, and {\ifmmode\ddot{O}\else\"{O}\fi}zel}}]{Guver2010Aug}
\bibinfo{author}{\bibfnamefont{T.}~\bibnamefont{G{\ifmmode\ddot{u}\else\"{u}\fi}ver}},
  \bibinfo{author}{\bibfnamefont{P.}~\bibnamefont{Wroblewski}},
  \bibinfo{author}{\bibfnamefont{L.}~\bibnamefont{Camarota}}, \bibnamefont{and}
  \bibinfo{author}{\bibfnamefont{F.}~\bibnamefont{{\ifmmode\ddot{O}\else\"{O}\fi}zel}},
  \bibinfo{journal}{Astrophys. J.} \textbf{\bibinfo{volume}{719}},
  \bibinfo{pages}{1807} (\bibinfo{year}{2010}{\natexlab{b}}), ISSN
  \bibinfo{issn}{0004-637X}.

\bibitem[{\citenamefont{{\ifmmode\ddot{O}\else\"{O}\fi}zel
  et~al.}(2009)\citenamefont{{\ifmmode\ddot{O}\else\"{O}\fi}zel,
  G{\ifmmode\ddot{u}\else\"{u}\fi}ver, and Psaltis}}]{Ozel2009Mar}
\bibinfo{author}{\bibfnamefont{F.}~\bibnamefont{{\ifmmode\ddot{O}\else\"{O}\fi}zel}},
  \bibinfo{author}{\bibfnamefont{T.}~\bibnamefont{G{\ifmmode\ddot{u}\else\"{u}\fi}ver}},
  \bibnamefont{and} \bibinfo{author}{\bibfnamefont{D.}~\bibnamefont{Psaltis}},
  \bibinfo{journal}{Astrophys. J.} \textbf{\bibinfo{volume}{693}},
  \bibinfo{pages}{1775} (\bibinfo{year}{2009}), ISSN \bibinfo{issn}{0004-637X}.

\bibitem[{\citenamefont{Elebert et~al.}(2009)\citenamefont{Elebert, Reynolds,
  Callanan, Hurley, Ramsay, Lewis, Russell, Nord, Kane, DePoy
  et~al.}}]{Elebert2009Apr}
\bibinfo{author}{\bibfnamefont{P.}~\bibnamefont{Elebert}},
  \bibinfo{author}{\bibfnamefont{M.~T.} \bibnamefont{Reynolds}},
  \bibinfo{author}{\bibfnamefont{P.~J.} \bibnamefont{Callanan}},
  \bibinfo{author}{\bibfnamefont{D.~J.} \bibnamefont{Hurley}},
  \bibinfo{author}{\bibfnamefont{G.}~\bibnamefont{Ramsay}},
  \bibinfo{author}{\bibfnamefont{F.}~\bibnamefont{Lewis}},
  \bibinfo{author}{\bibfnamefont{D.~M.} \bibnamefont{Russell}},
  \bibinfo{author}{\bibfnamefont{B.}~\bibnamefont{Nord}},
  \bibinfo{author}{\bibfnamefont{S.~R.} \bibnamefont{Kane}},
  \bibinfo{author}{\bibfnamefont{D.~L.} \bibnamefont{DePoy}},
  \bibnamefont{et~al.}, \bibinfo{journal}{Mon. Not. R. Astron. Soc.}
  \textbf{\bibinfo{volume}{395}}, \bibinfo{pages}{884} (\bibinfo{year}{2009}),
  ISSN \bibinfo{issn}{0035-8711}.

\bibitem[{\citenamefont{Abubekerov et~al.}(2008)\citenamefont{Abubekerov,
  Antokhina, Cherepashchuk, and Shimanskii}}]{Abubekerov2008May}
\bibinfo{author}{\bibfnamefont{M.~K.} \bibnamefont{Abubekerov}},
  \bibinfo{author}{\bibfnamefont{E.~A.} \bibnamefont{Antokhina}},
  \bibinfo{author}{\bibfnamefont{A.~M.} \bibnamefont{Cherepashchuk}},
  \bibnamefont{and} \bibinfo{author}{\bibfnamefont{V.~V.}
  \bibnamefont{Shimanskii}}, \bibinfo{journal}{Astron. Rep.}
  \textbf{\bibinfo{volume}{52}}, \bibinfo{pages}{379} (\bibinfo{year}{2008}),
  ISSN \bibinfo{issn}{1562-6881}.

\bibitem[{\citenamefont{Abbott et~al.}(2020)\citenamefont{Abbott, Abbott,
  Abraham, Acernese, Ackley, Adams, Adhikari, Adya, Affeldt, Agathos
  et~al.}}]{abbott/2020}
\bibinfo{author}{\bibfnamefont{R.}~\bibnamefont{Abbott}},
  \bibinfo{author}{\bibfnamefont{T.~D.} \bibnamefont{Abbott}},
  \bibinfo{author}{\bibfnamefont{S.}~\bibnamefont{Abraham}},
  \bibinfo{author}{\bibfnamefont{F.}~\bibnamefont{Acernese}},
  \bibinfo{author}{\bibfnamefont{K.}~\bibnamefont{Ackley}},
  \bibinfo{author}{\bibfnamefont{C.}~\bibnamefont{Adams}},
  \bibinfo{author}{\bibfnamefont{R.~X.} \bibnamefont{Adhikari}},
  \bibinfo{author}{\bibfnamefont{V.~B.} \bibnamefont{Adya}},
  \bibinfo{author}{\bibfnamefont{C.}~\bibnamefont{Affeldt}},
  \bibinfo{author}{\bibfnamefont{M.}~\bibnamefont{Agathos}},
  \bibnamefont{et~al.}, \bibinfo{journal}{Astrophys. J. Lett.}
  \textbf{\bibinfo{volume}{896}}, \bibinfo{pages}{L44} (\bibinfo{year}{2020}),
  ISSN \bibinfo{issn}{2041-8205}.

\bibitem[{\citenamefont{Most et~al.}(2018)\citenamefont{Most, Weih, Rezzolla,
  and Schaffner-Bielich}}]{Most2018Jun}
\bibinfo{author}{\bibfnamefont{E.~R.} \bibnamefont{Most}},
  \bibinfo{author}{\bibfnamefont{L.~R.} \bibnamefont{Weih}},
  \bibinfo{author}{\bibfnamefont{L.}~\bibnamefont{Rezzolla}}, \bibnamefont{and}
  \bibinfo{author}{\bibfnamefont{J.}~\bibnamefont{Schaffner-Bielich}},
  \bibinfo{journal}{Phys. Rev. Lett.} \textbf{\bibinfo{volume}{120}},
  \bibinfo{pages}{261103} (\bibinfo{year}{2018}), ISSN
  \bibinfo{issn}{1079-7114}.

\bibitem[{\citenamefont{Marino et~al.}(2018)\citenamefont{Marino, Degenaar,
  Di~Salvo, Wijnands, Burderi, and Iaria}}]{Marino2018Jun}
\bibinfo{author}{\bibfnamefont{A.}~\bibnamefont{Marino}},
  \bibinfo{author}{\bibfnamefont{N.}~\bibnamefont{Degenaar}},
  \bibinfo{author}{\bibfnamefont{T.}~\bibnamefont{Di~Salvo}},
  \bibinfo{author}{\bibfnamefont{R.}~\bibnamefont{Wijnands}},
  \bibinfo{author}{\bibfnamefont{L.}~\bibnamefont{Burderi}}, \bibnamefont{and}
  \bibinfo{author}{\bibfnamefont{R.}~\bibnamefont{Iaria}},
  \bibinfo{journal}{Mon. Not. R. Astron. Soc.} \textbf{\bibinfo{volume}{479}},
  \bibinfo{pages}{3634} (\bibinfo{year}{2018}), ISSN \bibinfo{issn}{0035-8711}.

\bibitem[{\citenamefont{Liu et~al.}(2014)\citenamefont{Liu, Zhang, and
  Wen}}]{Liu2014May}
\bibinfo{author}{\bibfnamefont{H.}~\bibnamefont{Liu}},
  \bibinfo{author}{\bibfnamefont{X.}~\bibnamefont{Zhang}}, \bibnamefont{and}
  \bibinfo{author}{\bibfnamefont{D.}~\bibnamefont{Wen}},
  \bibinfo{journal}{Phys. Rev. D} \textbf{\bibinfo{volume}{89}},
  \bibinfo{pages}{104043} (\bibinfo{year}{2014}), ISSN
  \bibinfo{issn}{2470-0029}.

\bibitem[{\citenamefont{Arba{\ifmmode\tilde{n}\else\~{n}\fi}il and
  Malheiro}(2015)}]{Arbanil2015Oct}
\bibinfo{author}{\bibfnamefont{J.~D.~V.}
  \bibnamefont{Arba{\ifmmode\tilde{n}\else\~{n}\fi}il}} \bibnamefont{and}
  \bibinfo{author}{\bibfnamefont{M.}~\bibnamefont{Malheiro}},
  \bibinfo{journal}{Phys. Rev. D} \textbf{\bibinfo{volume}{92}},
  \bibinfo{pages}{084009} (\bibinfo{year}{2015}), ISSN
  \bibinfo{issn}{2470-0029}.

\bibitem[{\citenamefont{Carvalho et~al.}(2018)\citenamefont{Carvalho,
  Arba{\ifmmode\tilde{n}\else\~{n}\fi}il, Marinho, and
  Malheiro}}]{Carvalho2018May}
\bibinfo{author}{\bibfnamefont{G.~A.} \bibnamefont{Carvalho}},
  \bibinfo{author}{\bibfnamefont{J.~D.~V.}
  \bibnamefont{Arba{\ifmmode\tilde{n}\else\~{n}\fi}il}},
  \bibinfo{author}{\bibfnamefont{R.~M.} \bibnamefont{Marinho}},
  \bibnamefont{and} \bibinfo{author}{\bibfnamefont{M.}~\bibnamefont{Malheiro}},
  \bibinfo{journal}{Eur. Phys. J. C} \textbf{\bibinfo{volume}{78}},
  \bibinfo{pages}{1} (\bibinfo{year}{2018}), ISSN \bibinfo{issn}{1434-6052}.

\bibitem[{\citenamefont{Rocha et~al.}(2020)\citenamefont{Rocha, Carvalho, Deb,
  and Malheiro}}]{rocha/2020}
\bibinfo{author}{\bibfnamefont{F.}~\bibnamefont{Rocha}},
  \bibinfo{author}{\bibfnamefont{G.~A.} \bibnamefont{Carvalho}},
  \bibinfo{author}{\bibfnamefont{D.}~\bibnamefont{Deb}}, \bibnamefont{and}
  \bibinfo{author}{\bibfnamefont{M.}~\bibnamefont{Malheiro}},
  \bibinfo{journal}{Phys. Rev. D} \textbf{\bibinfo{volume}{101}},
  \bibinfo{pages}{104008} (\bibinfo{year}{2020}),
  \urlprefix\url{https://link.aps.org/doi/10.1103/PhysRevD.101.104008}.

\bibitem[{\citenamefont{Banerjee et~al.}(2000)\citenamefont{Banerjee, Ghosh,
  and Raha}}]{Banerjee2000Jan}
\bibinfo{author}{\bibfnamefont{S.}~\bibnamefont{Banerjee}},
  \bibinfo{author}{\bibfnamefont{S.~K.} \bibnamefont{Ghosh}}, \bibnamefont{and}
  \bibinfo{author}{\bibfnamefont{S.}~\bibnamefont{Raha}}, \bibinfo{journal}{J.
  Phys. G: Nucl. Part. Phys.} \textbf{\bibinfo{volume}{26}},
  \bibinfo{pages}{L1} (\bibinfo{year}{2000}), ISSN \bibinfo{issn}{0954-3899}.

\bibitem[{\citenamefont{Cheng and Harko}(2000)}]{Cheng2000Sep}
\bibinfo{author}{\bibfnamefont{K.~S.} \bibnamefont{Cheng}} \bibnamefont{and}
  \bibinfo{author}{\bibfnamefont{T.}~\bibnamefont{Harko}},
  \bibinfo{journal}{Phys. Rev. D} \textbf{\bibinfo{volume}{62}},
  \bibinfo{pages}{083001} (\bibinfo{year}{2000}), ISSN
  \bibinfo{issn}{2470-0029}.

\bibitem[{\citenamefont{Harko and Cheng}(2002)}]{Harko2002Apr}
\bibinfo{author}{\bibfnamefont{T.}~\bibnamefont{Harko}} \bibnamefont{and}
  \bibinfo{author}{\bibfnamefont{K.~S.} \bibnamefont{Cheng}},
  \bibinfo{journal}{Astron. Astrophys.} \textbf{\bibinfo{volume}{385}},
  \bibinfo{pages}{947} (\bibinfo{year}{2002}), ISSN \bibinfo{issn}{0004-6361}.

\bibitem[{\citenamefont{{Morris} et~al.}(1988)\citenamefont{{Morris}, {Thorne},
  and {Yurtsever}}}]{morris/1988}
\bibinfo{author}{\bibfnamefont{M.~S.} \bibnamefont{{Morris}}},
  \bibinfo{author}{\bibfnamefont{K.~S.} \bibnamefont{{Thorne}}},
  \bibnamefont{and}
  \bibinfo{author}{\bibfnamefont{U.}~\bibnamefont{{Yurtsever}}},
  \bibinfo{journal}{\prl} \textbf{\bibinfo{volume}{61}}, \bibinfo{pages}{1446}
  (\bibinfo{year}{1988}).

\bibitem[{\citenamefont{Weber et~al.}(2012)\citenamefont{Weber, Orsaria,
  Rodrigues, and Yang}}]{Weber2012Aug}
\bibinfo{author}{\bibfnamefont{F.}~\bibnamefont{Weber}},
  \bibinfo{author}{\bibfnamefont{M.}~\bibnamefont{Orsaria}},
  \bibinfo{author}{\bibfnamefont{H.}~\bibnamefont{Rodrigues}},
  \bibnamefont{and} \bibinfo{author}{\bibfnamefont{S.-H.} \bibnamefont{Yang}},
  \bibinfo{journal}{Proc. Int. Astron. Union} \textbf{\bibinfo{volume}{8}},
  \bibinfo{pages}{61} (\bibinfo{year}{2012}), ISSN \bibinfo{issn}{1743-9213}.

\bibitem[{\citenamefont{Deb et~al.}(2018)\citenamefont{Deb, Rahaman, Ray, and
  Guha}}]{Deb2018Mar}
\bibinfo{author}{\bibfnamefont{D.}~\bibnamefont{Deb}},
  \bibinfo{author}{\bibfnamefont{F.}~\bibnamefont{Rahaman}},
  \bibinfo{author}{\bibfnamefont{S.}~\bibnamefont{Ray}}, \bibnamefont{and}
  \bibinfo{author}{\bibfnamefont{B.~K.} \bibnamefont{Guha}},
  \bibinfo{journal}{J. Cosmol. Astropart. Phys.}
  \textbf{\bibinfo{volume}{2018}}, \bibinfo{pages}{044} (\bibinfo{year}{2018}),
  ISSN \bibinfo{issn}{1475-7516}.

\bibitem[{\citenamefont{Carvalho
  et~al.}(2020{\natexlab{b}})\citenamefont{Carvalho, Dos~Santos, Moraes, and
  Malheiro}}]{Carvalho2020Jun}
\bibinfo{author}{\bibfnamefont{G.~A.} \bibnamefont{Carvalho}},
  \bibinfo{author}{\bibfnamefont{S.~I.} \bibnamefont{Dos~Santos}},
  \bibinfo{author}{\bibfnamefont{P.~H. R.~S.} \bibnamefont{Moraes}},
  \bibnamefont{and} \bibinfo{author}{\bibfnamefont{M.}~\bibnamefont{Malheiro}},
  \bibinfo{journal}{Int. J. Mod. Phys. D} \textbf{\bibinfo{volume}{29}},
  \bibinfo{pages}{2050075} (\bibinfo{year}{2020}{\natexlab{b}}), ISSN
  \bibinfo{issn}{0218-2718}.

\bibitem[{\citenamefont{Sharif and Waseem}(2018)}]{Sharif2018Oct}
\bibinfo{author}{\bibfnamefont{M.}~\bibnamefont{Sharif}} \bibnamefont{and}
  \bibinfo{author}{\bibfnamefont{A.}~\bibnamefont{Waseem}},
  \bibinfo{journal}{Eur. Phys. J. C} \textbf{\bibinfo{volume}{78}},
  \bibinfo{pages}{1} (\bibinfo{year}{2018}), ISSN \bibinfo{issn}{1434-6052}.

\bibitem[{\citenamefont{Majid and Sharif}(2020)}]{Majid2020Aug}
\bibinfo{author}{\bibfnamefont{A.}~\bibnamefont{Majid}} \bibnamefont{and}
  \bibinfo{author}{\bibfnamefont{M.}~\bibnamefont{Sharif}},
  \bibinfo{journal}{Universe} \textbf{\bibinfo{volume}{6}},
  \bibinfo{pages}{124} (\bibinfo{year}{2020}), ISSN \bibinfo{issn}{2218-1997}.

\bibitem[{\citenamefont{Maurya et~al.}(2019)\citenamefont{Maurya, Errehymy,
  Deb, Tello-Ortiz, and Daoud}}]{Maurya2019Aug}
\bibinfo{author}{\bibfnamefont{S.~K.} \bibnamefont{Maurya}},
  \bibinfo{author}{\bibfnamefont{A.}~\bibnamefont{Errehymy}},
  \bibinfo{author}{\bibfnamefont{D.}~\bibnamefont{Deb}},
  \bibinfo{author}{\bibfnamefont{F.}~\bibnamefont{Tello-Ortiz}},
  \bibnamefont{and} \bibinfo{author}{\bibfnamefont{M.}~\bibnamefont{Daoud}},
  \bibinfo{journal}{Phys. Rev. D} \textbf{\bibinfo{volume}{100}},
  \bibinfo{pages}{044014} (\bibinfo{year}{2019}), ISSN
  \bibinfo{issn}{2470-0029}.

\end{thebibliography}

\end{document}